\documentclass[a4paper,11pt]{article}
\usepackage{jheppub} 
\usepackage{lineno}
\usepackage{booktabs}
\usepackage{array}
\usepackage{caption}
\usepackage{multirow}
\usepackage{graphics}
\usepackage{cancel}
\usepackage{comment}
\usepackage{dcolumn}
\usepackage{bm}
\usepackage{soul}
\usepackage{subcaption}
\usepackage{enumitem}
\usepackage{xspace}
\usepackage{booktabs}
\usepackage{lipsum}
\usepackage[per-mode=power]{siunitx} 
\usepackage[ISO]{diffcoeff}
\usepackage[capitalise]{cleveref}
\usepackage{subcaption}
\usepackage{ragged2e}
\usepackage{mathrsfs}
\usepackage{cleveref}
\usepackage{hyperref}
\usepackage[normalem]{ulem}
\usepackage{amsmath}
\usepackage{wrapfig}
\usepackage{slashed}
\usepackage[dvipsnames]{xcolor}
\usepackage[compat=1.1.0]{tikz-feynman}

\DeclareSIUnit\year{yr} 
\def\arraystretch{2}\tabcolsep=10pt

\DeclareCaptionJustification{justified}{\justifying}
\newcommand{\mpl}{m_\mathrm{pl}}

\def\d{\mathrm{d}}

\def\C{\mathcal{C}}

\def\vec{\mathbf}

\def\e{\mathrm{e}}

\newcommand{\overbar}[1]{\mkern 1.5mu\overline{\mkern-1.5mu#1\mkern-1.5mu}\mkern 1.5mu}

\definecolor{rp}{cmyk}{0.2, 1, 0.6, 0}
\definecolor{rp}{cmyk}{0.2, 1, 0.6, 0}
\definecolor{green2}{cmyk}{0.27, 0, 1, 0.52}

\usepackage{color}
\hypersetup{
    colorlinks=true,       
    linkcolor=green2,          
    citecolor=green2,        
    filecolor=magenta,      
    urlcolor=green2           
}

\title{New insights into axion freeze-in}

\author{Mudit Jain, }
\emailAdd{mudit.jain@kcl.ac.uk}

\author{Angelo Maggi, }
\emailAdd{angelo.maggi@kcl.ac.uk}

\author{Wen-Yuan Ai,}
\emailAdd{wenyuan.ai@kcl.ac.uk}

\author{and David J.~E. Marsh}
\emailAdd{david.j.marsh@kcl.ac.uk}

\affiliation{Theoretical Particle Physics and Cosmology, King's College London, Strand, London, WC2R 2LS, United Kingdom}

\abstract{Freeze-in via the axion-photon coupling, $g_{\phi\gamma}$, can produce axions in the early Universe. At low reheating temperatures close to the minimum allowed value $T_{\rm reh}\approx T_{\rm BBN}\approx 10\,{\rm MeV}$, the abundance peaks for axion masses $m_\phi\approx T_{\rm reh}$. Such heavy axions are unstable and subsequently decay, leading to strong constraints on $g_{\phi\gamma}$ from astrophysics and cosmology. In this work, we revisit the computation of the freeze-in abundance and clarify important issues. We begin with a complete computation of the collision terms for the Primakoff process, electron-positron annihilation, and photon-to-axion (inverse-)decay, while approximately taking into account plasma screening and threshold effects. We then solve the Boltzmann equation for the full axion distribution function. We confirm previous results about the importance of both processes to the effective ``relic abundance'' (defined as density prior to decay), and provide useful fitting formulae to estimate the freeze-in abundance from the equilibrium interaction rate. For the distribution function, we find an out-of-equilibrium population of axions and introduce an effective temperature for them. We follow the evolution right up until decay, and find that the average axion kinetic energy is larger than a thermal relic by between 20\% and 80\%, which may have implications for limits on decaying axions from X-ray spectra. We extend our study to a two-axion system with quartic cross-coupling, and find that for typical/expected couplings, freeze-in of a second axion flavour by annihilations leads to a negligibly small contribution to the relic density.}

\begin{document}

\begin{flushright}
\raggedleft
    \small KCL-PH-TH/2024-31
\end{flushright}

\maketitle
\flushbottom


\section{Introduction}

Axion~\cite{Peccei:1977hh,Wilczek:1977pj,Weinberg:1977ma} and axion-like particles (ALPs) are pseudo-Goldstone bosons that naturally arise in theories beyond the Standard Model. They are often considered to be associated with some broken global chiral symmetries in the UV, which can also give rise to effective couplings with the Standard Model degrees of freedom, for instance, the Chern-Simons coupling to photons $g_{\phi\gamma} \phi F_{\mu\nu}\Tilde{F}^{\mu\nu}/4$, the derivative coupling to chiral fermionic currents $g_{\phi \psi}(\partial_\mu\phi) \bar{\psi} \gamma^\mu \gamma_5 \psi/(2m_\psi)$, etc. The same physics also arises in theories with extra dimensions (such as string theory), where there is no global chiral symmetry in four dimensions and the ALP arises as a pseudoscalar associated with higher dimensional gauge symmetries (see e.g. Ref.~\cite{Reece:2023czb}). Such couplings provide for an extremely rich phenomenology in cosmology, astrophysics, and particle physics (see reviews~\cite{Marsh:2015xka,DiLuzio:2020wdo,Marsh:2021jmi,Peccei:2006as,OHare:2024nmr,Sikivie:2006ni,Irastorza:2018dyq,Adams:2022pbo}).

Various mechanisms for generating ALPs are possible. These include the standard misalignment mechanism~\cite{Abbott:1982af,Preskill:1982cy,Dine:1982ah,Turner:1983he}, thermal freeze-out~\cite{Cadamuro:2010cz,Cadamuro:2011fd}, and more recently in scenarios involving low reheating temperatures, freeze-in~\cite{Hall:2009bx,Bernal:2017kxu,Arias:2023wyg}. Due to their interaction with photons, there are a wide variety of constraints on $g_{\phi\gamma}$~\cite{AxionLimits}. If ALPs come into thermal equilibrium, these constraints are so powerful as to effectively exclude ALPs with $m_a\gtrsim \mathcal{O}(\rm keV)$ with $g_{\phi\gamma}\gtrsim\alpha/(2\pi \mpl)$~\cite{Cadamuro:2011fd} (with $\alpha$ the fine-structure constant and $\mpl = 1/\sqrt{8\pi G}$ the reduced Planck mass). However, thermal equilibrium for ALPs relies on the assumption that the maximum thermalisation temperature of the Universe, $T_{\rm max}$, is sufficiently high. In particular, it needs to be higher than the freeze-out temperature of the Primakoff interaction~\cite{Cadamuro:2011fd}:
\begin{align}
    T_{\rm fo}\approx 1.2 \times 10^{6}\,\frac{\sqrt{g_\star(T_{\rm fo})}}{g_{\star,Q}(T_{\rm fo})}\left(\frac{10^{-11} {\rm GeV}^{-1}}{g_{\phi\gamma}}\right)^2 {\rm GeV}\,,
\end{align}
where $g_\star$ is the effective number of standard model relativistic degrees of freedom, and $g_{\star,Q}$ is the effective number of relativistic {\it charged} degrees of freedom. The latter is defined as $\sum_{i\in\{e,\mu\,...\}} Q_i^2 n_i(T)=\frac{3}{4} \frac{\zeta(3)}{\pi^2}g_{\star,Q} (T) T^3$. Here we have added the missing factor $3/4$ from Refs.~\cite{Cadamuro:2011fd,Depta:2020wmr} so that $g_{\star,Q}=4$ for one species of leptons or quarks in our definition. For example, in the Standard Model $g_{\star,Q}=4$ if $m_e<T \ll m_\mu$. As such, one must include the contributions from relativistic charged leptons and quarks, as the temperature gets comparable to, or above their masses.\footnote{We assume that the contributions from charged mesons and protons, for temperatures below the QCD phase transition, are negligible.} To ensure $T_{\rm fo}> T_{\rm BBN}$, one requires $g_{\phi\gamma}\lesssim 10^{-7}\,{\rm GeV}^{-1}$.

The success of Big Bang Nucleosynthesis (BBN) in predicting the light element abundances~\cite{ParticleDataGroup:2018ovx,Planck:2018vyg} is good evidence that the Universe was in a state of radiation domination and thermal equilibrium at temperatures $T_{\rm BBN}\sim \mathcal{O}(1-10)$ MeV, corresponding to a timescale of a few minutes. The thermal history of the Universe at times earlier than this is almost completely unconstrained. The epoch of initial conditions (inflation or otherwise) must end leaving the Universe dominated by particles or fields of some sort. The effective equation of state of the associated fluid, $w$, in the period between initial conditions and BBN, and whether or not the cosmic fluid was in any sort of thermal equilibrium, is uncertain. For this work, we define \emph{reheating} as the point in time from which the Universe was dominated by Standard Model degrees of freedom in thermal equilibrium, with its energy budget given by the Friedmann equation
\begin{align}
    \label{eq:Friedmann_H(T)}
    3H^2 \mpl^2 = \frac{\pi^2}{30}g_\star(T)T^4\,.
\end{align}
From the moment of reheating thus defined until matter radiation equality at redshift $z_{\rm eq}\approx 3400$~\cite{Planck:2018vyg}, the Universe is radiation dominated with no intervening epochs of matter domination or low-scale inflation. Under such an assumption the relic abundance is to be computed precisely. In modified cosmologies, such as those with early matter domination induced by coherent oscillations of moduli fields (e.g. Refs.~\cite{Coughlan:1983ci,Banks:1995dt,Acharya:2008bk,Iliesiu:2013rqa}), ``reheating'' thus defined corresponds to the decay of the last modulus. Reheating must occur at a temperature $T_{\rm reh}\geq T_{\rm BBN}$. Taking $T_{\rm reh} = T_{\rm BBN} \ll T_{\rm fo}$, and assuming, conservatively, that the epoch prior to reheating does not produce any axions (diluting any present before by inflation or entropy production) leading to a vacuum initial condition at $T_{\rm reh}$, the freeze-in process gives rise to an irreducible density of axions produced by interactions with the thermal bath at $T\leq T_{\rm reh}$~\cite{Langhoff:2022bij,Balazs:2022tjl}. This idea strongly motivates further detailed study of the freeze-in process at temperatures of $\mathcal{O}(1-10)$ MeV, which is the subject of this work.

A number of factors could complicate the physics of ALP freeze-in process and the resulting phenomenology, compared to the studies in Refs.~\cite{Langhoff:2022bij,Balazs:2022tjl}. For relatively large ALP masses (compared to the QCD axion), $m_a \sim$ MeV, and low reheating temperatures (close to BBN), ALPs freeze in and subsequently decay with phenomenological implications for BBN, the CMB, and X-rays, to name a few~\cite{Cadamuro:2011fd,Millea:2015qra,Depta:2020wmr,Langhoff:2022bij,Balazs:2022tjl}. Ref.~\cite{Depta:2020wmr} provides a method to compute the full distribution function of ALPs, while the recent phenomenology studied in Refs.~\cite{Langhoff:2022bij,Balazs:2022tjl} do not account for the non-equilibrium distribution, and compute the abundance prior to decay assuming stability. In the present work, we compute the full distribution function and study the evolution of the kinetic energy, which defines an effective temperature, right up until the point of decay. As we will show, ALPs produced by freeze-in have an effective temperature higher than that of a thermal relic, and we discuss possible implications for phenomenology. 

Another possible complication is that in theories with many ALPs, like string theory~\cite{Arvanitaki:2009fg,Svrcek:2006yi,Demirtas:2018akl}, there can be quartic (and possibly cubic) interactions among the different ALPs, while having hierarchically different couplings to photons~\cite{Gendler:2023kjt}. It is therefore possible that one ALP species, $\phi_1$, could be abundantly produced by freeze-in, and then in turn produce a second, electromagnetically inert species, $\phi_2$, by 2-2 annihilations or decays. The case of decays was considered briefly in Ref.~\cite{Gendler:2023kjt}. In this work, we study the case of 2-2 annihilations by an interaction $\lambda\phi_1^2\phi_2^2/4$.

The structure of the paper is as follows.  In Sec.~\ref{sec2}, we delve into the freeze-in production of ALPs by examining all of the possible tree-level processes. For the leading processes, which are the Primakoff reaction and photon to axion (inverse-)decays, we trace their complete evolution until ALPs decay into photons. To this end, we comprehensively compute the relevant collision terms and solve the Boltzmann equation. Additionally, we provide straightforward fitting formulae derived upon comparison with the corresponding equilibrium relic abundance. Finally, we extend our investigation to a two-ALP system. In Sec.~\ref{sec3}, we explore non-thermal features of the ALP distribution function. We introduce an effective temperature based on the averaged kinetic energy, and compare it with that of thermal relics of equivalent mass (assumed to have decoupled at the reheating temperature). Discussion and conclusions are presented in Sec.~\ref{sec4}. Technical details are provided in the appendices. Appendix~\ref{appendixA} offers a detailed computation of the collision terms, while Appendix~\ref{app:numerical_evolvefk} discusses the numerical method for the evolution of the distribution function. In Appendix~\ref{app:reheating_postdecays}, we briefly explore the heating effects in the primordial plasma, resulting from ALP decays into photons.

Throughout, we work in natural units, $\hbar = c = k_{\rm B} = 1$, and adopt the mostly negative metric signature $(+,-,-,-)$.

\section{Freeze-in of ALPs in the early Universe}
\label{sec2}
\subsection{ALPs from the Standard Model Plasma}

\begin{figure}[ht]
    \centering
    \includegraphics[scale=0.5]{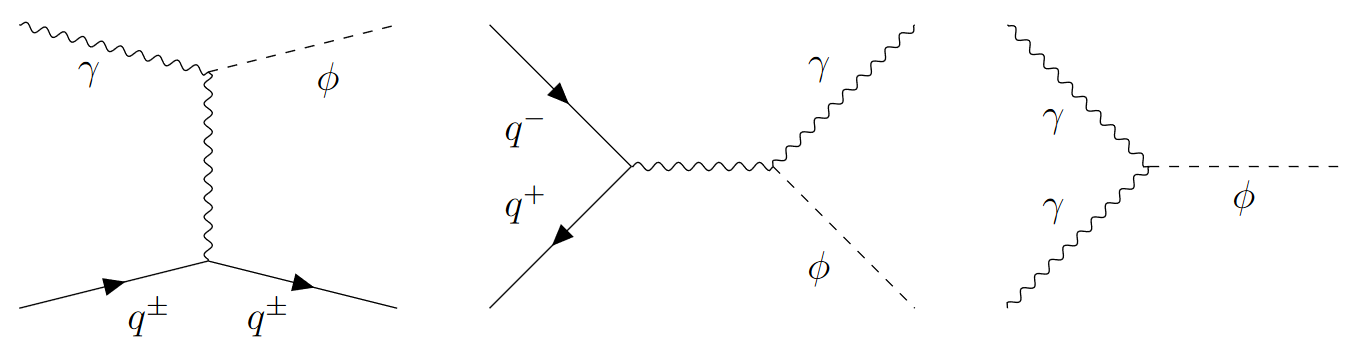}
    \caption{Feynman diagrams for the three processes for ALP production, considered in this work: the Primakoff process (leftmost), the fermion-anti-fermion annihilation process (middle), and the photon-to-ALP inverse decay (rightmost). With the annihilation reaction being sub-dominant, only the Primakoff and inverse/forward decays are of primary importance.}
    \label{fig:Feyn}
\end{figure}

We begin by investigating the production of ALPs (of mass $m_{\phi}$) in the early Universe, facilitated by the Chern-Simons vertex 
\begin{align}
    \mathcal{L}_{\rm \phi\gamma} = - \frac{g_{\phi\gamma}}{4} \phi F_{\mu\nu}\tilde{F}^{\mu\nu}\,.
\end{align}
At the leading order, this interaction gives rise to three distinct processes (see Fig.~\ref{fig:Feyn}): the Primakoff reaction involving electromagnetically charged fermions, $q^{\pm}\gamma \rightarrow q^{\pm} \phi$, the fermion anti-fermion annihilation reaction $q^{+}q^{-}\rightarrow \phi\gamma$, and the (inverse and forward) decay between photons and ALPs, $\phi \rightleftharpoons \gamma\gamma$.

The corresponding Boltzmann equation for the distribution function (occupation number function) $f^{\phi}_{\bm k}$ of ALPs is linear and can be written as follows:
\begin{align}
\label{eq:fboltzman}
    \frac{\partial f^{\phi}_{k}}{\partial t} - Hk\frac{\partial f^{\phi}_{k}}{\partial k} &= \Bigl[(1+f^{\phi}_{k})\,\mathcal{C}_{\rm Prim}(k) - f^{\phi}_{k}\,\mathcal{C}'_{\rm Prim}(k)\Bigr] + \Bigl[(f^{\phi,{\rm eq}}_{k} - f^{\phi}_{k})\,\mathcal{C}_{\rm decay}(k)\Bigr]\nonumber\\
    &\, + \Bigl[(1+f^{\phi}_{k})\,\mathcal{C}_{\rm ann}(k) - f^{\phi}_{k}\,\mathcal{C}'_{\rm ann}(k)\Bigr]\,,
\end{align}
where $k=|{\bm k}|$.\footnote{In the appendices, we use non-bold letters, $k$, $p$ etc., to denote four-momenta. We hope that difference between the modulus of three momenta and the four-momenta is clear from the context.} Here, the collision terms $\mathcal{C}_{\rm Prim}$ and $\mathcal{C}'_{\rm Prim}$ correspond to the forward and backward Primakoff reactions, $\mathcal{C}_{\rm decay}$
corresponds to photon-ALP decays, and
$\mathcal{C}_{\rm ann}$ and $\mathcal{C}'_{\rm ann}$ correspond to the forward and backward annihilation reactions. In this work, we focus on the low reheating temperature scenario of the early Universe, in which case none of the reactions ever come into equilibrium. That is, the associated rates are much smaller than the Hubble parameter. In such a scenario, ALPs are produced via the freeze-in mechanism~\cite{Depta:2020wmr,Langhoff:2022bij,Balazs:2022tjl}. While both the Primakoff and annihilation collision terms can be simplified under the assumption of small occupation numbers of photons and electrons, i.e. neglecting Bose enhancement and Pauli blocking, together with negligible ALP (and fermion) masses~\cite{Bolz:2000fu,Langhoff:2022bij} (also see~\cref{sec:Primakoff_rel_calculation} and ~\cref{sec:annihilation_rel_calculation} for the relevant calculations), we note that in the scenario of low reheating temperatures and for ALP masses not so small as compared to the reheating temperature, we cannot adhere to such approximations. We evaluate the Primakoff collision term semi-analytically, details of which are given in~\cref{app:PartPrim_rate} (see Eq.~\eqref{eq:collision_PartPrim} in particular for $\mathcal{C}_{\rm Prim}$ and $\mathcal{C}'_{\rm Prim}$). On the other hand for the decay process, a full analytical expression can be obtained. See~\cref{app:dec_rate} for details (in particular Eq.~\eqref{eq:decayrate} for $\mathcal{C}_{\rm decay}$). Also, while the above equation applies in general, for our scenario of ALP freeze-in where $f^{\phi}_k \ll 1$, we can neglect the $\mathcal{C}'_{\rm Prim}$ and $\mathcal{C}'_{\rm ann}$ terms and also set $1+f^{\phi}_{k}$ to unity. We evolve the Boltzmann equation~\eqref{eq:fboltzman} for the ALP distribution function numerically (see~\cref{app:numerical_evolvefk} for relevant details).

We also note that while we will have appropriately accounted for the ALP mass dependence, we adhere to approximations to only minimally capture some plasma effects, as usually done in the literature~\cite{Cadamuro:2010cz}. Specifically, we insert the plasma frequency $m_\gamma (T)$ (c.f. Eq.~\eqref{eq:photonmass}) in the photon propagator for the Primakoff process in order to capture screening effects; and we assume massive photon (of mass $m_\gamma(T)$) for the inverse decay in order to account for threshold effects. Finite temperature effects would modify the two-point functions of particles in the plasma. Not only would this change their propagators, but also modify their dispersion relations, especially for the transverse and longitudinal modes of photons. Although we do not expect that such finite temperature effects would significantly affect the results presented in this paper, we will improve upon these approximations in an upcoming work.

\subsubsection{Full relic abundance}

For a concrete illustration of the ALP freeze-in mechanism, we shall fix the reheating temperature near the lower bound $T_{\rm reh} = 10\,\mathrm{MeV}$.\footnote{See Refs.~\cite{Hasegawa:2019jsa,Kawasaki:2000en} for a concrete lower bound of $T_{\rm reh} \geq 4-5$ MeV.} For such low reheating temperatures, only the electrons and positrons contribute in the Primakoff reaction, since muons are much more massive than $T_{\rm reh}$. For larger values of $T_{\rm reh}$ (but still smaller than that needed for the Primakoff reaction to coming into equilibrium), our analysis can be extended to include more charged species (e.g. if $T_{\rm reh}$ becomes comparable to or larger than $m_{\mu} \approx 105.6$ MeV, then we can include contributions from muons and anti-muons as well). Also, since the scaling with $g_{\phi\gamma}$ trivially goes as $g_{\phi\gamma}^2$ for the different reaction channels considered, we shall fix the ALP-photon coupling constant $g_{\phi\gamma} = 10^{-11}\,\mathrm{GeV}^{-1}$, while only varying the mass $m_{\phi}$. Considering three prototypical masses $m_{\phi} = \{0.1, 1, 10\}$ MeV, let us now discuss the ALP freeze-in mechanism in some detail. Throughout our presentation, we shall use inverse temperature $1/T$ as a proxy for time.\\

\begin{figure}[ht]
  \centering  \includegraphics[width=0.75\linewidth]{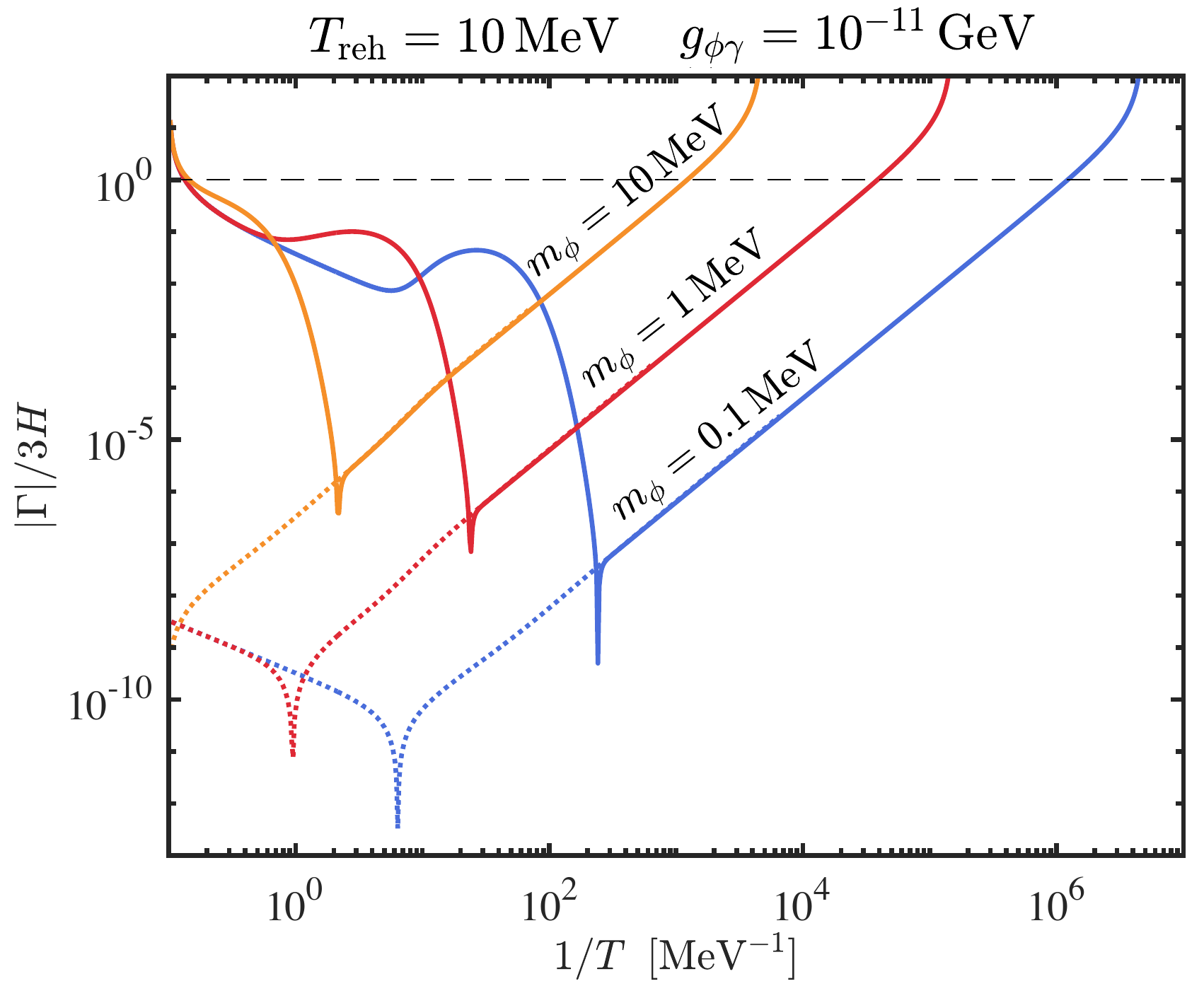}
\caption{The ratio of the absolute valued total rate $|\Gamma_{\rm all}|$ over $3H$, due to all the relevant processes (in solid). Here we also show the curves for the absolute valued equilibrium rate $|\Gamma_{\rm eq}|$ (dashed), to draw distinctions and comparisons.}
\label{fig:Gammaover3H}
\end{figure}

Let us begin by first analyzing the reaction rates due to all the processes. For this purpose, we define the following effective rate due to all the reactions: 
\begin{align}
\label{eq:Gamma_all}
    \Gamma_{\rm all} \equiv \frac{1}{n}\int\frac{{\rm d}^3 {\bm k}}{(2\pi)^3}\,[{\rm RHS\,of\,Eq.~\eqref{eq:fboltzman}}]\,,
\end{align}
where $n \equiv (2\pi)^{-3}\int{\rm d}^3{\bm k}\,f^{\phi}_{k}$ is the ALP number density. The modulus of this, with respect to the Hubble rate, is shown in Fig.~\ref{fig:Gammaover3H} (solid curves). Firstly, we have verified that within our approximations the electron-positron annihilation reaction is always subdominant compared with the Primakoff reaction (c.f. Ref.~\cite{Langhoff:2022bij}; also see~\ref{app:annihilation_rate}). Therefore we shall only focus on the Primakoff and inverse/forward decays below. Starting with the Primakoff reaction, whose rate is maximum at the reheating temperature, we get the production of a primary population of ALPs. While this reaction rate (as compared to Hubble) keeps on decreasing, as soon as $m_{\gamma}(T) < m_{\phi}/2$ inverse decays start to occur (notice the change in behaviour of solid curves --- appearance of the first saddle point). This leads to a secondary population of ALPs, production of which is maximized around $m_{\gamma}(T) \simeq m_{\phi}/4$ (see the discussion around Eq.~\eqref{eq:Cdecay_leading}). However, this channel too becomes Boltzmann suppressed as the temperature of the plasma falls below $m_{\phi}$ (reflected by the exponential fall off of solid curves -- the presence of $f^{\phi,\rm eq}_k$ in the second bracket of the right-hand side of Eq.~\eqref{eq:fboltzman}). This epoch marks the final freeze-in of the ALPs. After this point the forward decay rate (which is negative), while being still much smaller than the Hubble rate, takes over. This is reflected by the ``spikes" in the figure which are caused by a sign change in $\Gamma_{\rm all}$. As $H$ keeps on decreasing, $|\Gamma_{\rm all}|/H$ keeps on increasing and ultimately when the ratio becomes order unity, ALPs decay away into photons. 

It may come as a surprise that $|\Gamma_{\rm all}|/3H$ is not that small, and even starts off being larger than unity in the beginning. However, we note that in the conventional comparison with the Hubble parameter, one uses the equilibrium decay rate $\Gamma_{\rm eq} \equiv n_{\rm eq}\langle\sigma v\rangle_{\rm eq}$. To illustrate the point better, in Fig.~\ref{fig:Gammaover3H} we have also shown the $|\Gamma_{\rm eq}|/3H$ curves (in dashed), obtained using the forward Primakoff reaction and the ALPs to photons forward decays (with the annihilation reaction neglected):
\begin{align}
\label{eq:Gamma_eq}
    \Gamma_{\rm eq} \equiv \frac{1}{n_{\rm eq}}\int\frac{\mathrm{d}^3{\bm k}}{(2\pi)^3}\,[(1+f^{\phi,\rm eq}_{k})\,\mathcal{C}_{\rm Prim}(k) - f^{\phi,\rm eq}_{k}\,\mathcal{C}_{\rm decay}(k)]\,.
\end{align}
As we can see in the figure, $|\Gamma_{\rm eq}|/3H$ is much smaller than unity from the beginning, indicating that the Primakoff reaction never comes into thermal equilibrium. This may also be understood from the fact that the period for $|\Gamma_{\rm all}|/3H>1$ at the beginning is too short so there is not enough time to reach equilibrium. Under the assumption of $f^{\phi,\rm eq}_k \ll 1$, the contribution towards the numerator of both $\Gamma_{\rm eq}$ and $\Gamma_{\rm all}$ from the Primakoff process, is the same. The difference only comes from the denominator, i.e. whether it is the equilibrium number density $n_{\rm eq}$, or the actual number density $n \ll n_{\rm eq}$ (which starts out from zero), which is the reason why the solid and dashed curves are disparate up until the ALPs to photons forward decay rate comes to dominate. For phenomenological purposes, later we shall provide a quantitative formula for the number density of ALPs from the Primakoff and inverse decay processes, using the -- simpler to use -- equilibrium rate $\Gamma_{\rm eq}$.\\

One of the most important quantities of interest is the (decaying) ALP DM fraction:
\begin{equation}
\label{eq:xi_def}
    \xi(T) = \frac{\rho(T)}{\bar{\rho}}\frac{s_0}{s(T)} \frac{1}{\Omega_{\rm DM}h^2}\,.
\end{equation}
Here $\rho(T) = (2\pi)^{-3}\int{\rm d}^3{\bm k}\,\omega_{k}\,f^{\phi}_{k}(T)$ is the energy density of the ALPs, which at late times (when the ALPs become non-relativistic) becomes mass times the number density, $\rho \rightarrow m_{\phi}n$. Also, $s(T)$ is the physical entropy density of the SM plasma with $s_0 \approx 2.9\times 10^9\,{\rm m}^{-3}$ its present-day value. The ratio $s_0/s(T)$ is just the volume redshift factor $a(T)^3/a_0^3$. The normalising density $\bar{\rho} = 1.06 \, \times 10^4\,{\rm MeV}\, {\rm m}^{-3}$, and $\Omega_{\rm DM}h^2 \approx 0.12$ is the relic dark matter fraction today. We will focus on the parameter space when $\xi (1+z_{\rm eq})/(1+z) \ll 1$, so that the standard cosmology holds to leading order. With this, let us now discuss the contributions from the Primakoff and inverse decay reactions in some detail, along with a quantitative formula for their respective contributions towards total freeze-in abundance.\\

\begin{figure}[ht]
  \centering\includegraphics[width=1.0\linewidth]{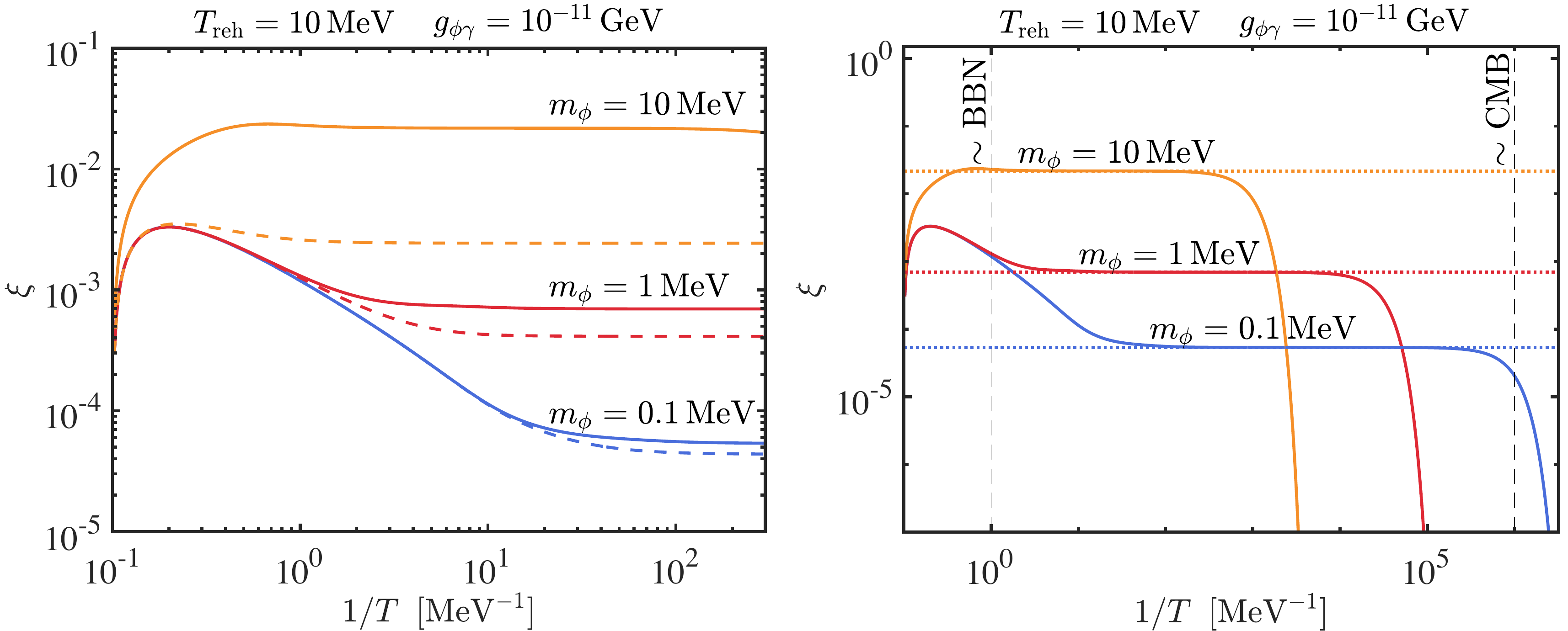}
\caption{Left panel: ALP DM fraction $\xi$ as it develops over time, for our three prototypical ALP masses. In dashed is shown the contribution from Primakoff only, while solid represents the total abundance, i.e. including the contribution from photon to ALP conversions. Right panel: Same as the left panel but with an extended time range to capture ALP to photon eventual decay.}
\label{fig:xiplot}
\end{figure}

\subsubsection{Understanding and fitting the relic abundance}
\noindent{\bf\uline{Primakoff process}:} To analyze the freeze-in abundance of ALPs due to the Primakoff reaction only, we set $\mathcal{C}_{\rm decay} = 0 = \mathcal{C}_{\rm ann}$ in Eq.~\eqref{eq:fboltzman}.\footnote{As mentioned earlier, the contribution from annihilations ($\mathcal{C}_{\rm ann}$) is sub-dominant and hardly makes any difference in our results. In particular, see Fig.~\ref{fig:annihilation_vs_Primakoff} in~\cref{app:annihilation_rate}.} See dashed curves in the left panel of Fig.~\ref{fig:xiplot}. Leaving a detailed calculation of the collision term aside for Appendix~\ref{app:PartPrim_rate}, let us sketch a qualitative understanding of the relevant physics. The Primakoff reaction is maximum at the reheating temperature and is only ``active" until the temperature of the SM plasma drops below the threshold energy required to produce ALPs. In particular for $m_{\phi} \gtrsim m_e$, the threshold is set by $T \sim m_\phi$, while for $m_{\phi} \lesssim m_e$, the threshold becomes $T \sim m_e$. This is because below these threshold temperatures, the Primakoff reaction starts to shut off: For the former case, this is due to there not being enough energy in the plasma to create ALPs of higher masses than $T$, while for the latter case, this is due to electrons and positrons themselves becoming Boltzmann suppressed.

\begin{figure}[h]
  \centering  \includegraphics[width=1\linewidth]{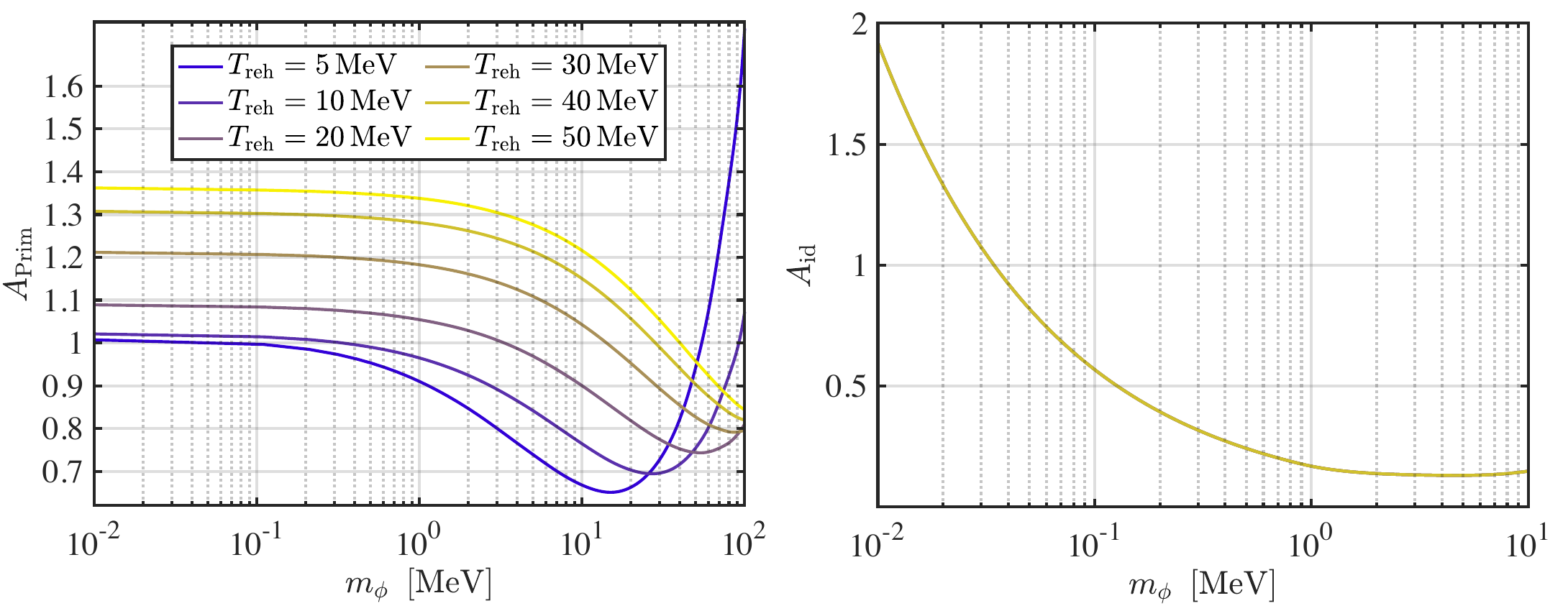}
\caption{Left panel: The fitting coefficient $A_{\rm Prim}$, for the number density of ALPs from the Primakoff process. Right panel: The fitting coefficient $A_{\rm id}$, for the number density of ALPs from photon to ALPs inverse decays. In both panels, we show six curves for reheating temperature $T_{\rm reh} = \{5, 10, 15, 20, 25, 30\}$ MeV (in increasing order of the colour darkness). Notice that $A_{\rm id}$ is independent of the reheating temperature when $2.5 m_{\phi} \lesssim T_{\rm reh}$ (where $2.5 m_{\phi} \approx \overbar{T}_{\rm id}$).}
\label{fig:Aid_n_APrim}
\end{figure}

While a general expression for the Primakoff-produced ALP DM fraction is difficult to obtain, we can fit the late time freeze-in number density of ALPs by using the equilibrium number density, $n_{\rm eq}$, multiplied by the ratio of the equilibrium reaction rate over Hubble, $\Gamma_{\rm eq}/H$, as done in Refs.~\cite{Hall:2009bx,Jaeckel:2014qea,Gendler:2023kjt}. Since the reaction rate is maximized at $T_{\rm reh}$ (see Fig.~\ref{fig:Gammaover3H}), we can therefore write~\cite{Jaeckel:2014qea}
\begin{align}
\label{eq:neff_Primakoff}
    n_{\rm Prim}(T \ll T_{\rm reh}) = A_{\rm Prim}\left(\frac{a(T_{\rm reh})}{a(T)}\right)^3\Biggl[n_{\rm eq}\frac{\Gamma_{\rm eq}}{H}\Biggr]_{T = T_{\rm reh}}\,,
\end{align}
where $A_{\rm Prim} = A_{\rm Prim}(m_{\phi},T_{\rm reh})$ is a fitting factor. Collecting all of the $m_{\phi}$ dependence in $A_{\rm Prim}$ and $n_{\rm eq}$, we can further use the simplified rate expression when the ALP and electron mass are neglected~\cite{Cadamuro:2011fd}:
\begin{align}
\label{eq:Gamma_Prim_eq}
    \Gamma_{\rm eq} \approx \frac{\alpha g^2_{\phi\gamma}T^3}{12}\Bigl[2\log\left(\frac{T}{m_{\gamma}}\right) + 0.82\Bigr]\,.
\end{align}
From the plasma mass $m_\gamma = (e^2n_e/\langle\omega_e\rangle)^{1/2}$ where $n_e$ and $\langle\omega_e\rangle$ are the equilibrium number density and average energy of electrons (positrons), we have $m_{\gamma} \simeq eT/3 \simeq 0.1T$ for $T_{\rm reh} \gg  m_e$. Using this along with the usual Hubble rate~\eqref{eq:Friedmann_H(T)}, we get
\begin{align}
\label{eq:neff_Primakoff_final}
    n_{\rm Prim}(T \ll T_{\rm reh}) = A_{\rm Prim}\,\frac{2.4 \times 10^{-8}}{\sqrt{g_{\star}(T_{\rm reh})}}\left(\frac{g_{\phi\gamma}}{10^{-11}\,{\rm GeV}^{-1}}\right)^2\left(\frac{T_{\rm reh}}{10\,{\rm MeV}}\right)\left(\frac{a(T_{\rm reh})}{a(T)}\right)^3n_{\rm eq}(T_{\rm reh})\,.
\end{align}
See the left panel of Fig.~\ref{fig:Aid_n_APrim} for $A_{\rm Prim}$ as a function of ALP mass $m_{\phi}$ for 5 different reheating temperatures.\footnote{In Appendix~\ref{app:TableAPrim}, we also provide a table of values for $A_{\rm Prim}$ for different axion masses and reheat temperatures upto $50$ MeV (when the contribution from muons becomes $\mathcal{O}(10\%)$).} Furthermore for simplicity, one can use the relativistic approximation $n_{\rm eq}(T_{\rm reh}) \approx \zeta(3)T_{\rm reh}^3/\pi^2$ for $m_{\phi} \ll T_{\rm reh}$ in the above expression. 

Before analyzing photons to ALPs inverse decays, let us briefly discuss some of the features of this fitting factor $A_{\rm Prim}$: (1) The changing asymptotic values of $A_{\rm Prim}$ towards small $m_{\phi}$, for different reheating temperatures $T_{\rm reh}$, is due to the changing number of relativistic degrees of freedom (both $g_s$ and $g_{\star}$). The fact that the Primakoff process is not instantaneous and that $g_s$ and $g_{\star}$ change during the production of ALPs, results in a slightly different dependence of $n_{\rm Prim}$ on $g_{\star}$ and $g_{s}$ than what is dictated by the right-hand side of the above parameterization; (2) As $m_{\phi}$ increases, the decreasing feature (followed by subsequent rise) is due to the complicated mass dependence of $\mathcal{C}_{\rm Prim}$ on $m_{\phi}$, which is not captured in $\Gamma_{\rm eq}$. It is important to note that only when $m_{\phi} \lesssim \mathcal{O}(0.01) \times T_{\rm reh}$, does the neglection of mass becomes justified, and the simplified rate expression~\eqref{eq:Gamma_Prim_eq} can be used to calculate the late time abundance $n_{\rm Prim}$.\\

\noindent{\bf\uline{Photon-ALPs inverse decays}:} Next, we include the decay processes, i.e. switching $\mathcal{C}_{\rm decay}$ in Eq.~\eqref{eq:fboltzman} back on. The total ALP freeze-in abundance, due to both Primakoff and decays, is shown in solid in the left panel of Fig.~\ref{fig:xiplot}. With a detailed calculation of the rate given in~\cref{app:dec_rate}, here we outline some of the relevant aspects and also derive an effective formula for the freeze-in abundance due to photon to ALP inverse decays. To begin, note that this process is kinematically blocked until the plasma frequency (which can be thought of as the effective mass for the photons) drops below half of the ALP's mass, i.e. $2m_{\gamma}(T) < m_{\phi}$, which then defines the threshold temperature $T_{\rm th}$. As soon as this happens, photons in the plasma can combine to produce ALPs. However, once $T \lesssim m_\phi$, inverse decays get suppressed since there is not enough energy in the plasma anymore to produce more ALPs. (In fact, it becomes exponentially suppressed, as also reflected from the $f^{\phi,\rm eq}_{k}$ term in Eq.~\eqref{eq:fboltzman}). With this, let us now estimate when is the inverse decay most efficient. For this purpose, it is natural to work with the two cases of $T_{\rm th} \gtrsim m_{e}$ and $T_{\rm th} \lesssim m_{e}$, since during this transition the dependence of photon mass on the plasma temperature changes drastically.\\

Starting with the $T_{\rm th} \gtrsim m_{e}$ case, we focus on the non-relativistic modes $k/m_{\phi} \rightarrow 0$ since they give the dominant contribution. The collision term $\mathcal{C}_{\rm decay}$ takes the following form (c.f. Eq.~\eqref{eq:decayrate})
\begin{align}
\label{eq:Cdecay_leading}
    \mathcal{C}_{\rm decay} = \frac{g^2_{\phi\gamma}m^3_{\phi}}{8\pi z^3}(z^3-1)^{3/2}\coth\left(z\frac{m_{\gamma}}{2T}\right) + \mathcal{O}(k^2) \equiv \Gamma_{\rm id}\,,
\end{align}
where $z \equiv m_{\phi}/2m_{\gamma}$. It maximizes near $z = \bar{z}_{\rm id} = 2$, i.e. $m_{\phi} = 4m_{\gamma}(\overbar{T}_{\rm id})$, giving $\Gamma_{\rm id}(T=\overbar{T}_{\rm id}) \approx 0.09g^2_{\phi\gamma}m^3_{\phi}\coth(m_{\phi}/4\overbar{T}_{\rm id})$. Now if the ALP mass is sufficiently larger than the reheating temperature such that $\overbar{T}_{\rm id} \gtrsim T_{\rm reh}$, then we cannot evaluate the above term at $\overbar{T}_{\rm id}$. In this case, we should only evaluate it at $T_{\rm reh}$. With this, we define the following formula to capture the freeze-in abundance of ALPs due to inverse decays:
\begin{align}
\label{eq:neff_invdecays}
    n_{\rm id}(T \ll T') &\equiv A_{\rm id}\Biggl[\left(\frac{a(T')}{a(T)}\right)^3n_{\rm eq}(T')\frac{\Gamma_{\rm id}(T')}{H(T')}\Biggr]_{T' = {\rm min}[\overbar{T}_{\rm id}\,, T_{\rm reh}]}\,,
\end{align}
where $A_{\rm id} = A_{\rm id}(m_{\phi})$ is once again a phenomenological factor that fixes the equality. We will confirm that $A_{\rm id}$ does not depend on $T_{\rm reh}$.

Recall that for $T\gtrsim m_e$, the plasma frequency becomes $m_{\gamma} \simeq eT/3 \simeq 0.1T$, giving $T_{\rm th}\simeq 5 m_{\phi}$ and 
\begin{align}
    \overbar{T}_{\rm id} \simeq 2.5 m_{\phi}\,.
\end{align}
Also, the condition $T_{\rm th}\gtrsim m_e$ gives $m_{\phi} \gtrsim 0.2m_e \simeq 0.1$ MeV. Using this together with the usual Hubble rate from Eq.~\eqref{eq:Friedmann_H(T)}, we get
\begin{align}
    n_{\rm id}(T \ll T')\Bigr|_{m_{\phi} \gtrsim 0.2m_e} &\approx A_{\rm id}\times 10^{-8}\left(\frac{g_{\phi\gamma}}{10^{-11}\,{\rm GeV}^{-1}}\right)^2\left(\frac{m_{\phi}}{0.1\,{\rm MeV}}\right)\Biggl[\left(\frac{a(T')}{a(T)}\right)^3\frac{n_{\rm eq}(T')}{\sqrt{g_{\star}(T')}}\nonumber\\
    &\times\Biggl\{\frac{T'}{\overbar{T}_{\rm id}}\frac{\left(8\left(\frac{\overbar{T}_{\rm id}}{T'}\right)^3-1\right)^{3/2}}{7^{3/2}}\frac{\coth\left(0.1\frac{\overbar{T}_{\rm id}}{T'}\right)}{\coth(0.1)}\Biggr\}\Biggr]_{T' = {\rm min}[\overbar{T}_{\rm id}\,, T_{\rm reh}]}\,.
\end{align}
In the above expression, one can further use the relativistic approximation $n_{\rm eq} \approx \zeta(3)T^3/\pi^2$ (with the error compared to the actual $n_{\rm eq}$ being always less than about $\sim 7.6\%$). For $\overbar{T}_{\rm id} < T_{\rm reh}$, i.e. ${\rm min}[\overbar{T}_{\rm id}\,, T_{\rm reh}] = \overbar{T}_{\rm id}$, the factor in the parenthesis above becomes unity.\\
 
For the case $T_{\rm th}\lesssim m_e$ which implies $m_\phi \lesssim 0.2 m_e \approx 0.1$ MeV, the plasma frequency becomes $m_{\gamma} \approx e(n_e/m_e)^{1/2}$ with $n_e \approx (m_eT/2\pi)^{3/2}e^{-m_e/T}$ being the non-relativistic thermal distribution. This drops exponentially with $T$, giving a non-trivial dependence of the threshold temperature on $m_{\phi}$, $m_{\gamma}(T_{\rm th}) = m_{\phi}/2$. While in this case, we expect that the produced ALPs are relativistic, we still work with the non-relativistic formula Eq.~\eqref{eq:Cdecay_leading}, leaving all the non-trivial $m_{\phi}$ dependence to $A_{\rm id}$. Using $m_{\gamma}(\bar{T}_{\rm id}) = m_{\phi}/4$, we get 
\begin{align}
    \overbar{T}_{\rm id} \approx 0.34\Biggl[{\rm PLog}\Bigl(1.17\Bigl(\frac{0.1\,{\rm MeV}}{m_{\phi}}\Bigr)^{4/3}\Bigr)\Biggr]^{-1}\,{\rm MeV}\,,
\end{align}
where PLog is the ProductLog function. Since $m_{\phi} \lesssim 0.2 m_e$, we can Taylor expand the PLog function for $m_{\phi} \ll 0.2m_{e}$ to get $\overbar{T}_{\rm id} \approx 0.58[\log(0.1\,{\rm MeV}/m_{\phi})]^{-1}$. While this is a nice simplification, it is only applicable for $m_{\phi} \lesssim 0.01$ MeV since only then the percentage error (compared to the PLog function) is less than order ten. Nevertheless, using $\overbar{T}_{\rm id}$ (together with $z = 2$) in Eq.~\eqref{eq:Cdecay_leading}, and expanding the $\coth$ function for $\overbar{T}_{\rm id} \gg m_{\phi}$, we get the following from Eq.~\eqref{eq:neff_invdecays}:
\begin{align}
    n_{\rm id}(T \ll \overbar{T}_{\rm id})\Bigr|_{m_{\phi} \lesssim 0.2m_e} &\approx \frac{A_{\rm id}\times 10^{-8}}{\sqrt{g_{\star}(\overbar{T}_{\rm id})}}\left(\frac{g_{\phi\gamma}}{10^{-11}\,{\rm GeV}}\right)^2\left(\frac{m_{\phi}}{0.1\,{\rm MeV}}\right)^2\left(\frac{0.26\,{\rm MeV}}{\overbar{T}_{\rm id}}\right)\times\nonumber\\
    &\qquad\qquad\qquad\qquad\qquad\qquad\qquad\qquad \left(\frac{a(\bar{t}_{\rm id})}{a(t)}\right)^3 n_{\rm eq}(\overbar{T}_{\rm id})\,.
\end{align}

While the above captures the contribution of photon to ALPs inverse decays for $m_{\phi} \lesssim 0.2m_{e} \approx 0.1$ MeV, it is to be noted that in general this contribution becomes sub-dominant compared with the Primakoff contribution (c.f. Eq.~\eqref{eq:neff_Primakoff_final}), as $m_{\phi}$ becomes smaller. Ultimately when the $e^- + e^+ \leftrightarrow 2\gamma$ reaction decouples and electrons freeze out, i.e. around $T_{ee} \approx 16$ keV~\cite{Thomas:2019ran}, the plasma mass $m_{\gamma}(T_{ee})$ saturates to $\approx 0.95$ meV. 
This means that for ALP masses $m_{\phi} < 1.9$ meV, photon to ALP inverse decays effectively never occur in the early Universe plasma.

See the right panel of Fig.~\ref{fig:Aid_n_APrim} for $A_{\rm id}$ as a function of ALP mass $m_{\phi}$. As expected, $A_{\rm id}$ does not depend on the reheating temperature for the relevant case of $m_{\phi} \lesssim T_{\rm reh}$ (while for higher masses it becomes exponentially difficult to produce ALPs). We find that for $m_{\phi} \lesssim 1$ MeV, $A_{\rm id}$ follows a power law, while for larger masses it can be roughly approximated by a constant (with a relative error less than $\sim \mathcal{O}(10)\%$):
\begin{align}
    A_{\rm id} \approx 0.16\left(\frac{m_{\phi}}{1 \,{\rm MeV}}\right)^{-0.53}\, \Theta(m_{\phi}\leq 1 \rm MeV) + 0.16 \, \Theta(m_{\phi}>1 \rm MeV)\,.
\end{align}
 \\

\noindent{\bf\uline{Total late time abundance}:} Using the above two separate analyses, the total number density of the frozen in ALPs is the sum of the two:
\begin{align}
    n(T) = n_{\rm Prim}(T \ll T_{\rm reh}) + n_{\rm id}(T \ll T')\Bigr|_{T' = {\rm min}[\overbar{T}_{\rm id}\,, T_{\rm reh}]}\,,
\end{align}
with the contribution from electron positron annihilations being negligible. Using this, the late time fractional abundance when ALPs become non-relativistic, i.e. $\rho \rightarrow m_{\phi}n$ in Eq.~\eqref{eq:xi_def}, and until they decay away into photons, becomes
\begin{align}
\label{eq:xi_latetime}
    \xi(T) = \frac{m_{\phi}n(T)}{\bar{\rho}}\frac{s_0}{s(T)} \frac{1}{\Omega_{\rm DM}h^2}\,.
\end{align}\\

\noindent{\bf\uline{ALPs-Photons forward decays}:} Finally, the forward axion to photon decay rate, while completely negligible during the freeze-in process, keeps on climbing up relatively to the Hubble parameter (reflected by the straight curves in Fig.~\ref{fig:Gammaover3H}). Eventually around the time when $\Gamma_{\rm fd} = g^2_{\phi\gamma}m^3_{\phi}/64\pi \approx H$, ALPs decay away into photons. Using Eq.~\eqref{eq:Friedmann_H(T)}, one obtains\footnote{For $m_\phi\lesssim 0.1\text{ MeV}$ decays happen after matter-radiation equality and this expression should be modified to include the change in the Friedmann equation.}
\begin{align}
\label{eq:T_forwarddecay}
    T_{\rm fd} \simeq \frac{60}{g^{1/4}_{\star}(T_{\rm fd})}\left(\frac{m_\phi}{1\,{\rm MeV}}\right)^{3/2}\left(\frac{g_{\phi\gamma}}{10^{-11}\,{\rm GeV}^{-1}}\right)\,{\rm eV}\,.
\end{align}
This is the same as the ``re-equilibration" temperature~\cite{Millea:2015qra,Depta:2020wmr}. With this, if the plasma was still tightly coupled (i.e. epoch prior to recombination), ALPs to photon decays would result in its temperature rise, otherwise leading to a secondary population of photons. We calculate and briefly discuss this heating in Appendix~\ref{app:reheating_postdecays}.

\subsection{ALPs from ALP annihilations}

Let us now consider the case of multiple interacting ALPs, motivated by string theory~\cite{Svrcek:2006yi,Arvanitaki:2009fg,Acharya:2010zx,Cicoli:2012sz}. At the effective theory level, we consider  for simplicity two ALPs interacting through the following quartic coupling 
\begin{align}
    \mathcal{L} =\frac{1}{2}\sum_{i=1,2}\left[\partial_\mu\phi_i\partial^\mu\phi_i- m_{\phi_i}^2\phi_i^2\right] -\frac{g_{\phi\gamma}}{4}\phi_1 F_{\mu\nu}\Tilde{F}^{\mu\nu} -\frac{\lambda}{4} \phi_1^2\phi^2_2\,,
\end{align}
where we have assumed that there is no coupling between the second ALP $\phi_2$ and photons. Motivated by string axiverse, we write
\begin{align}
    \lambda=\frac{m_{\phi_2}^2}{\mathcal{F}^2}\,,
\end{align}
where $\mathcal{F}$ is independent of $g_{\phi\gamma}$ in general.

At the leading order, the $\phi_2$ particles can be produced only from the annihilation $\phi_1\phi_1\rightarrow \phi_2\phi_2$. Its distribution function is governed by 
\begin{align}
    \frac{\partial f_k^{\phi_2}}{\partial t}-H k\frac{\partial f_k^{\phi_2}}{\partial k}= C_\lambda(k)\,,
\end{align}
where $C_\lambda$ is the collision term for the quartic interaction and is discussed in Appendix~\ref{ClambdaAppendix}. The corresponding DM fraction for the secondary ALP is
\begin{align}
    \xi_{\lambda}(T) \equiv \frac{m_{\phi_2}n_{\phi_2}(T)}{\bar{\rho}}\frac{s_0}{s(T)} \frac{1}{\Omega_{\rm DM}h^2}\,,
\end{align}
where $n_{\phi_2}\equiv \int\frac{\d^3{\bm k}}{(2\pi)^3} f^{\phi_2}_k$ is its number density.

\begin{figure}[ht]
\centering
\includegraphics[width=1\linewidth]{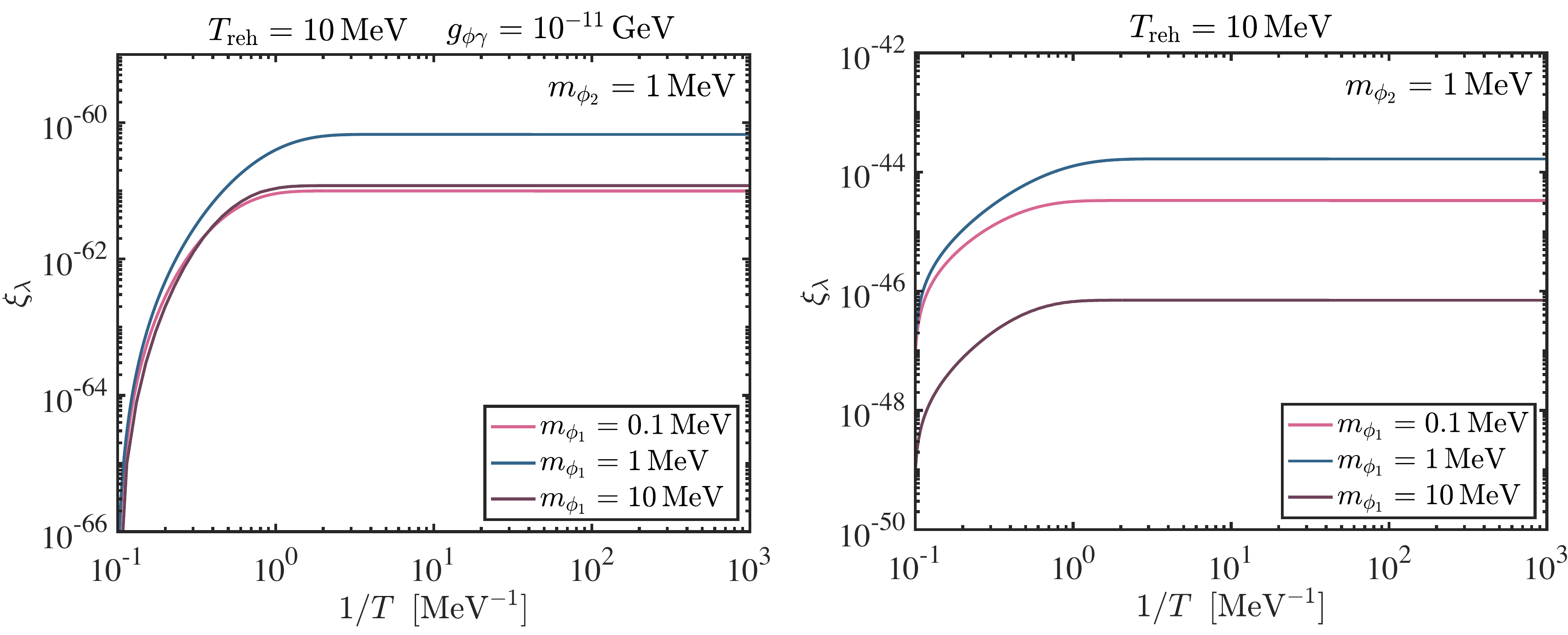}
\caption{ Left panel: The DM fraction $\xi_\lambda$ for the axion $\phi_2$ of mass $m_{\phi_2} = 1$ MeV, produced via $\phi_1\phi_1 \rightarrow \phi_2\phi_2$ quartic reaction. For each of the three cases of $m_{\phi_1} = \{0.1, 1, 10\}$ MeV, the axion $\phi_1$ was produced via the Primakoff and inverse decays (as discussed previously) with $g_{\phi\gamma} = 10^{-11}$ GeV$^{-1}$, while the secondary axion $\phi_2$ is produced solely via the quartic interaction with $\mathcal{F} = 10^{13}$ GeV (giving $\lambda = m_{\phi_2}^2/\mathcal{F}^2 =  10^{-32}$). Right panel: Same as left panel, but with the primary axion (axion $\phi_1$) being a thermal relic that also froze out around $T_{\rm reh} = 10$ MeV. The secondary ALP abundance is much larger in this case, as expected.}
\label{fig:xi_forlambda}
\end{figure}

In Fig.~\ref{fig:xi_forlambda}, we show the DM fraction of $\phi_2$ particles, from the $\phi_1$ annihilation ($\phi_1\phi_1\rightarrow \phi_2\phi_2$). We fix the $\phi_2$ mass at $1$ MeV, while consider three different masses of $\phi_1$ ($m_{\phi_1}=\{0.1,1,10\}$ MeV). The other couplings are chosen as $g_{\phi\gamma}=10^{-11}$ GeV$^{-1}$ and $\mathcal{F}=10^{13}$ GeV, which gives $\lambda=10^{-32}$. Also, we take $T_{\rm reh} = 10$ MeV as before. In the left panel, the primary ALPs ($\phi_1$) are produced via the freeze-in through the Primakoff and inverse decay processes as discussed previously. For comparison, in the right panel, we show the same quantity $\xi_\lambda$ assuming $\phi_1$ to be freeze-out thermal relics (with the added assumption that their temperature is equal to that of the Standard Model plasma initially, i.e. $T_{\rm relic} = T_{\rm reh}$). As expected, the secondary ALPs $\phi_2$ are frozen-in on account of the $2$-$2$ annihilation. Although the DM fraction of $\phi_2$ is negligible for both cases (on account of the very small coupling constant $\lambda$), one can see a significant enhancement for the case when $\phi_1$ is a freeze-out thermal relic, compared with the freeze-in discussed in this work. This is because of the much larger number density of the primary ALPs in the thermal relic case. Interestingly, there is no monotonic relation between $\xi_\lambda$ and $m_{\phi_1}$. 

For the $\phi_2$ DM fraction in the second case, i.e., when $\phi_1$ relic abundance is generated from freeze-out, one may give a quantitative estimate as follows. The number density of $\phi_2$ may be estimated as
\begin{align}
    n_{\phi_2}(T\ll T_{\rm reh}) \sim \left[\frac{\Gamma_{\lambda} }{H} n_{\phi_1}\right]_{T=T_{\rm reh}}\,,
\end{align}
where the interaction rate for the $2\rightarrow 2$ process $\Gamma_\lambda$ is given by
\begin{align}
    \Gamma_{\lambda} \sim \langle n_{\phi_1} \sigma v\rangle \sim \frac{\lambda^2 n_{\phi_1}}{16\pi \bar{s}}\,,
\end{align}
with $\bar{s}$ being the averaged Mandelstam $s$-variable, that can be estimated using the average energy per boson $\bar{s}=(2\langle \omega\rangle)^2$. Assuming a relativistic Boltzmann distribution for $\phi_1$ particles, we have $n_{\phi_1}(T)\sim \zeta(3) T^3/\pi^2$, and then $\langle \omega\rangle=(2\pi)^{-3}\int \d^3{\bm k}\,\omega_k f_k \approx 2.7\, T$. Together with $s(T)=2\pi^2 g_s\, T^3/45$ and the Friedmann equation~\eqref{eq:Friedmann_H(T)}, we arrive at\footnote{We have used the quoted numbers for $s_0$ and $\bar{\rho}$ below Eq.~\eqref{eq:xi_def}, and $\Omega_{\rm DM}h^2\approx 0.12$.}
\begin{align}
    \xi_\lambda(T\ll T_{\rm reh})\sim 10^{-43}\left(\frac{\lambda}{10^{-32}}\right)^2\left(\frac{10.54}{g_s}\right)\left(\sqrt{\frac{10.54}{g_\star}}\right)\left(\frac{m_{\phi_2}}{1\, {\rm MeV}}\right)\left(\frac{10\, {\rm MeV}}{T_{\rm reh}}\right)\,,
\end{align}
which roughly agrees with the magnitude of the numerical results in the right panel of Fig.~\ref{fig:xi_forlambda}. It is to be noted that the relic number density $n_{\phi_1}$ is not a relativistic Boltzmann distribution, and in the above rough estimate we do not expect to get accurate scaling with $m_{\phi_1}$.

Above, we have chosen $\mathcal{F}=10^{13}$ GeV so that $\lambda$ is strongly suppressed. In such a case, one can imagine that the abundance of $\phi_2$ would be negligible even without giving a detailed analysis. However, type IIB String Axiverse allows low-scale decay constants. For instance, in the case of isotropic compactification of the Calabi-Yau manifold, the axion decay constants are related to its volume, ${\rm V_{\rm CY}}$, as~\cite{Baumann:2014nda} 
\begin{align}
    \mathcal{F}\sim \frac{m_{\rm pl}}{\mathcal{V}^{1/3}}\,,
\end{align}
where $\mathcal{V}=V_{\rm CY}\times M_s^6$ and $M_s$ is the string scale. To have computational perturbative control, one usually requires that $\mathcal{V} \gg 1$. We can require $M_s \gtrsim 10$ TeV from collider experiments, while at the same time $V_{\rm CY}^{1/6} \lesssim 1\,\mu{\rm m}$ from tests of gravity. This, in principle, gives plenty of room to have small values of $\mathcal{F}$. For instance with $\mathcal{F} \sim 10$ TeV and $m_{\phi} = 1$ MeV, we get $\lambda\sim 10^{-14}$ implying $\xi_\lambda\sim 10^{-7}$. This could become relevant for cosmology.

One can also extend the analysis to other processes. For example, consider the $1\rightarrow 3$ process through the coupling $\mathcal{L}\supset -g \phi_1\phi_2^3/3!$. Then we can estimate
\begin{align}
    \langle n_{\phi_1}\sigma v\rangle &=\frac{1}{n_{\phi_1}^{(0)}}\int\frac{\d^3 {\bm k}_1 }{(2\pi)^3 2\omega_{{\bm k}_1}^{\phi_1}  }\int\frac{\d^3 {\bm k}_2 }{(2\pi)^3 2\omega_{{\bm k}_2}^{\phi_2}  }\int\frac{\d^3 {\bm k}_3 }{(2\pi)^3 2\omega_{{\bm k}_3}^{\phi_2}  }\int\frac{\d^3 {\bm k}_4 }{(2\pi)^3 2\omega_{{\bm k}_4}^{\phi_2}  }\times \e^{-\omega_{{\bm k}_1}/T}\notag\\
    &\quad\times (2\pi)^4 \delta^4(k_1-k_2-k_3-k_4) |\mathcal{M}|^2\notag\\
    &\sim \frac{g^2}{ 64\pi^3 m_{\phi_1} } \int_{\Omega_1} \d \omega^{\phi_2}_{{\bm k}_3} \int_{\Omega_2} \d \omega^{\phi_2}_{{\bm k}_4}\,.
 \end{align}
Above $\Omega_1$ and $\Omega_2$ are constraints enforced by kinematics. For a rough estimate, we approximate the integral as
\begin{align}
    \langle n_{\phi_1}\sigma v\rangle &\sim \frac{g^2}{64\pi^3 m_{\phi_1} } \int_{m_{\phi_2}}^{m_{\phi_1}-2m_{\phi_2}}\d \omega^{\phi_2}_{{\bm k}_3} \int_{m_{\phi_2}}^{m_{\phi_1}-m_{\phi_2}-\omega^{\phi_2}_{{\bm k}_3}  } \d \omega^{\phi_2}_{{\bm k}_4 }\notag\\
    &=\frac{g^2 (m_{\phi_1}- 3m_{\phi_2})^2 }{128\pi^3  m_{\phi_1}}\theta(m_{\phi_1}-3m_{\phi_2})\,.
\end{align}
In the last step, we have put the condition $m_{\phi_1}\geq 3 m_{\phi_2}$ by hand. Following the previous analysis for the $2\rightarrow 2$ process, we obtain
\begin{align}
    \xi_g (T\ll T_{\rm reh})&\sim \theta(m_{\phi_1}-3m_{\phi_2})\times 3\times 10^{-45}\notag\\
    &\times  \left(\frac{g}{10^{-32}}\right)^2\left(\frac{m_{\phi_2}/m_{\phi_1}}{0.1}\right)\left(\frac{m_{\phi_1}-3m_{\phi_2}}{1\, \rm MeV}\right)^2\left(\frac{10.54}{g_s}\right)\left(\sqrt{\frac{10.54}{g_\star}}\right) \left(\frac{10\,{\rm MeV}}{T_{\rm reh}}\right)^2\,.
\end{align}

\section{Non-thermal distribution of ALP momenta}
\label{sec3}

Since the ALPs are produced via freeze-in under the scenario of low reheating temperatures, it is obvious that their distributions are not the thermal equilibrium (Boltzmann) ones. In this section, we discuss this in some detail and also extract an effective temperature of these ALPs relative to the corresponding thermal relics (which decoupled at $T_{\rm reh}$ and have the same masses as ALPs). This can be useful for phenomenological purposes and for placing robust constraints on the ALP parameter space.\\

\begin{figure}[h]
  \centering  \includegraphics[width=1\linewidth]{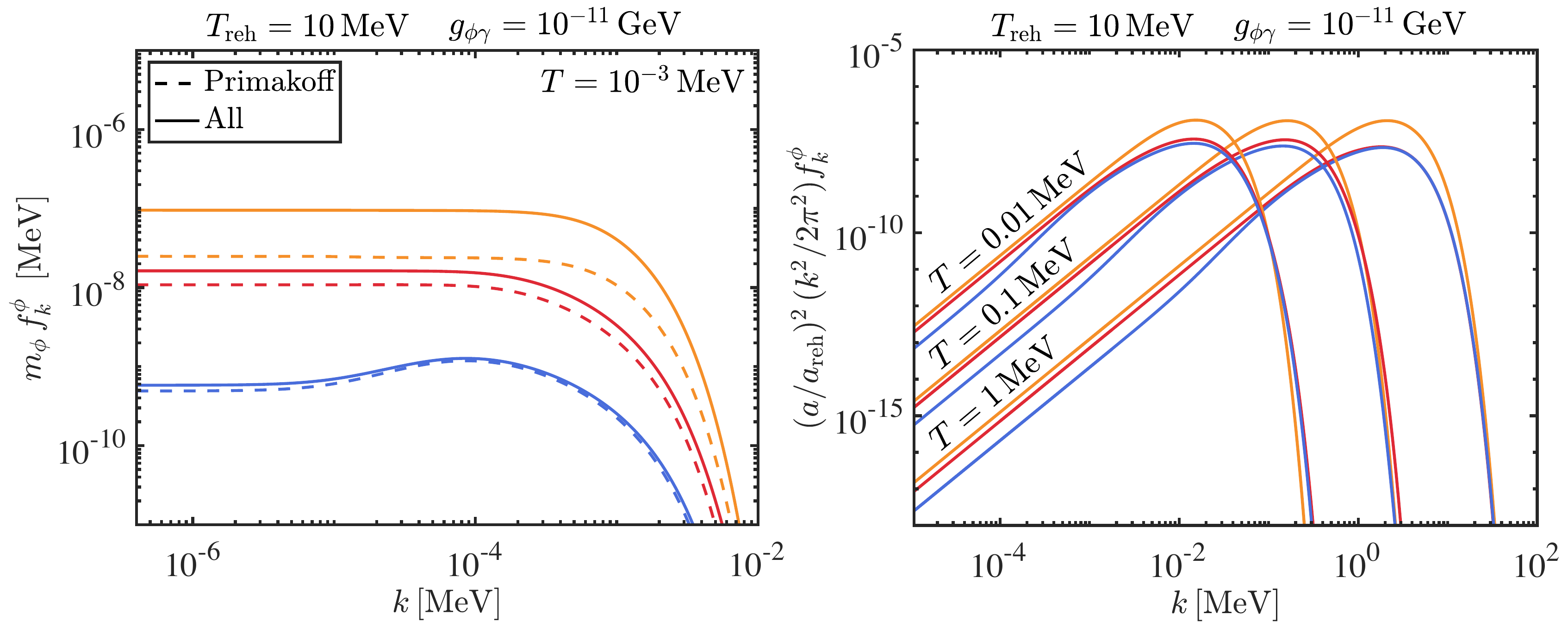}
\caption{Left panel: (Non-thermal) distribution function $f^{\phi}_{k}$ for the three prototypical ALPs (rescaled by their respective mass for better visual clarity), at $T = 10^{-3}$ MeV. The dashed curves are from the Primakoff process alone, while the solid curves are when both Primakoff and inverse decays are included. Right panel: (Non-thermal) power spectrum for the three prototypical ALPs, at three different times (corresponding to $T = [0.01, 0.1, 1]$ MeV). For a better comparison, here we have rescaled the spectrum by the factor $a^2(T)/a^2(T_{\rm reh})$ at these three times, in order to kill the red-shifting factor arising from the $k^2$ scaling. The power spectrum peaks at the typical momentum.
}
\label{fig:fk_all_bothpanels}
\end{figure}

In Fig.~\ref{fig:fk_all_bothpanels}, we present the distribution of ALPs for our three prototypical masses, while also separating out the contribution from the Primakoff process only. In particular, note the ``bump" feature in the distribution function $f_{k}$ for small masses, arising due to the inclusion of plasma frequency $m_{\gamma}$ in the photon's propagator (in the Primakoff reaction rate). Since $m_{\gamma} \approx 0.1T$ and most of the Primakoff production happens at $T_{\rm reh}$, the physical position of the peak is at $k \sim 0.1T_{\rm reh}a_{\rm reh}/a(t)$. This, of course, is relevant only for small ALP masses such that $m_{\phi} \lesssim m_{\gamma}(T_{\rm reh}) \approx 0.1T_{\rm reh}$. 

\begin{figure}[ht]
  \centering  \includegraphics[width=1\linewidth]{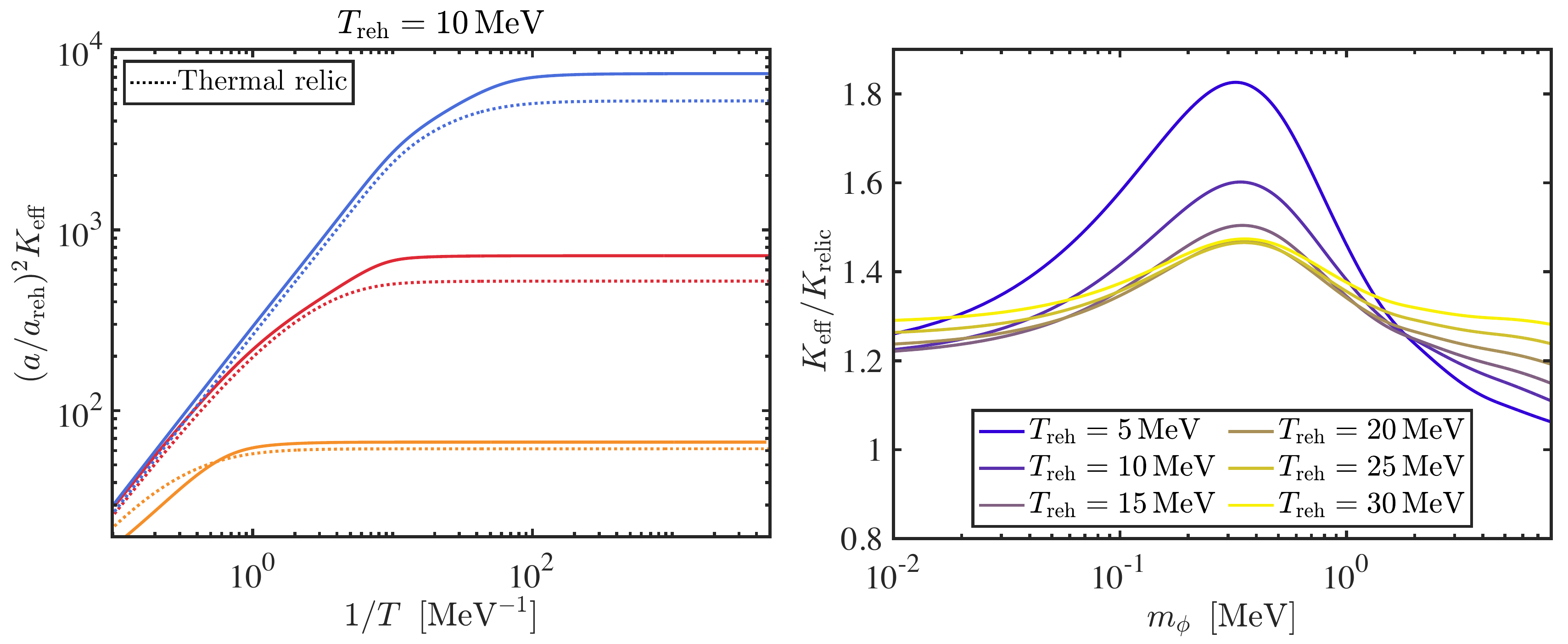}
\caption{Left panel: Effective kinetic energy $K_{\rm eff} \equiv \langle E_{\phi}\rangle - m_{\phi}$, rescaled by square of the scale factor $a^2(T)/a^2(T_i)$. Also shown in dotted are the equivalent curves for a warm thermal relic of the same mass. The ALPs are hotter than their equivalent thermal relics (reflective of the non-thermal distribution of ALPs). Right panel: Ratio of kinetic energies, i.e. effective temperatures, for the ALPs and corresponding thermal relics. See text for a discussion on the shape. In general, ALPs are hotter than the corresponding thermal relics.}
\label{fig:keff_bothpanels}
\end{figure}

To capture an ``effective temperature" for the axions, we calculate the following average kinetic energy
\begin{align}
    K_{\rm eff} \equiv \langle\omega_{k}\rangle - m_a\,,
\end{align}
where the bra-ket represents averaging using the full (non-thermal) ALP distribution $f^{\phi}_{k}$. For the three prototypical masses, this is shown in the left panel of Fig~\ref{fig:keff_bothpanels}. Here we have rescaled the effective kinetic energy with the square of the scale factor, in order to better see the transition towards the non-relativistic regime when $K_{\rm eff}(a/a_{\rm reh})^2 \rightarrow {\rm const}$. We also compare this with an equivalent warm thermal relic of the same mass, namely, a thermal relic of mass $m_a$ which decoupled from the plasma at $T_{\rm reh}$. The right panel of Fig.~\ref{fig:keff_bothpanels} shows the ratio of $K_{\rm eff}/K_{\rm relic}\gtrsim 1$, indicating that in general the axion is hotter than the corresponding thermal relic.

In general, we find that the ALPs are hotter than thermal relics, and can be even hotter by about $80\%$ depending on $m_a$ and $T_{\rm reh}$. While a full quantitative understanding of this ratio for different $T_{\rm reh}$ is quite involved, we provide a qualitative description of the overall shape and most importantly the appearance of the peak. Let us begin by mentioning that the ALPs produced via the Primakoff process alone can be up to $\sim 30\%$ warmer than the corresponding thermal relics,\footnote{We find $\sim 40\%$ warmer ALPs for $T_{\rm reh} \approx 50$ MeV. For larger reheating temperatures when muons cannot be neglected, we expect an even warmer population of ALPs due to the Primakoff reaction.} with this fraction roughly being independent of $m_{\phi}$. The peak arises solely due to the photons to ALPs inverse decays, and in particular due to the different behaviour of $m_{\gamma}(T)$ before and after $T \sim m_e$. Recall that the inverse decays mostly happen around $\overbar{T}_{\rm id} \simeq 2.5 m_{\phi}$, giving $m_{\phi} \simeq 0.4 m_{e} \simeq 0.2$ MeV for when $T \sim m_e$. For smaller ALP masses, $\overbar{T}_{\rm id}$ changes only logarithmically with $m_{\phi}$ since $m_{\gamma}$ decreases exponentially like $\sim T^{1/2}(m_eT)^{1/4}\e^{-m_e/2T}$, resulting in the production of dominantly relativistic ALPs which are much warmer than a corresponding thermal relic.\footnote{We find that for such lower mass ALPs, their temperature (solely due to inverse decays) is about $3$ times higher than corresponding thermal relics. However, this does not show up in $K_{\rm eff}$ since at these masses the Primakoff dominates the overall distribution function, and the average covers both channels.} Therefore as $m_{\phi}$ decreases and becomes smaller than $\sim 0.2$ MeV, ALPs are much warmer than due to just the Primakoff reaction alone, and the ratio $K_{\rm eff}/K_{\rm relic}$ rises. However, as $m_{\phi}$ decreases further, the total abundance of ALPs produced via inverse decays itself decreases and becomes only subdominant as compared to that from the Primakoff reaction. Therefore the ratio decreases down to values dictated by the latter.

The fact that the ALPs produced from freeze-in are hotter than thermal relics, may have some consequences for the dark matter distribution in the Milky Way. A warm dark matter (WDM) component will have a position-dependent DM fraction (defined as $\mathcal{F}_{\rm WDM}=\rho_{\rm WDM}/(\rho_{\rm WDM}+\rho_{\rm CDM})$)~\cite{Anderhalden:2012qt,Langhoff:2022bij}
\begin{align}
    \mathcal{F}^{\rm MW}_{\rm WDM}(r)=\mathcal{F}_{\rm WDM}\left[1+\zeta\cdot \left(\frac{\rm keV}{m_{\rm WDM}}\right)^2\left(\frac{\rm kpc}{r}\right) \right]^{-1}\,,
\end{align}
where $\zeta=0.008$ is a factor fitted by numerical results~\cite{Anderhalden:2012qt}. We anticipate that the result of $20-80$ $\%$ hotter axions can be accommodated in the above formula by replacing $m_a$ with $\mathcal{O}(1.2-1.8)\times\,m_a$, in effect making $\zeta$ smaller by $\mathcal{O}(1.2^2-1.8^2)$. A more careful analysis of how the bounds are affected is beyond the scope of this work, and must be addressed in a future study.

\section{Discussion and conclusions}
\label{sec4}

Axions and ALPs arise abundantly in theories of physics beyond the Standard Model, and also offer an exceptionally rich phenomenology, especially due to their chiral couplings with photons and fermions. The chiral Chern-Simons ALP-photon coupling opens avenues for detecting axions in laboratories~\cite{Graham:2013gfa,Kahn:2016aff,Caldwell:2016dcw,ADMX:2018gho,Ouellet:2018beu,Ouellet:2019tlz,DMRadio:2022jfv}, or through astrophysical objects (e.g.~\cite{Carroll:1989vb,Harari:1992ea,Plascencia:2017kca,Ivanov:2018byi,Davoudiasl:2019nlo,Liu:2019brz,Fedderke:2019ajk,Caputo:2019tms,Chen:2019fsq,Yuan:2020xui}). This coupling leads to three significant particle reactions: ALP forward decay, inverse decay, and the Primakoff process. These processes play crucial roles in the production and evolution of ALPs in the early Universe and can impose stringent constraints on the parameter space through various astrophysical phenomena, such as from the effects of axion decays on BBN and recombination.

In this work, we have conducted a thorough examination of these processes, accompanied by a full computation of the corresponding collision terms. While our discussion on the latter is general, our focus has been on the freeze-in scenario for ALP production, in contexts of low reheating temperatures. Contrary to conventional freeze-out scenarios, this scenario can open up parameter regions that are usually thought to be excluded. However, in the freeze-in scenario, solving the full integro-differential Boltzmann equation becomes necessary, complicating the analysis due to the non-thermal axion distribution. Consequently, we have investigated generic features concerning the entire axion evolution and a crucial quantity in axion cosmology, namely the ALP decaying DM fraction $\xi$ (cf.~\eqref{eq:xi_def}), with the aim of offering fresh perspectives. The primary outcomes are summarized below.

We began by introducing the effective interaction rate, $\Gamma_{\rm all}$ (cf.~\eqref{eq:Gamma_all}), obtained from integrating all collision terms in the Boltzmann equation over the momentum space. We investigated the ALP dark matter fraction originating from the Primakoff process, electron positron annihilations, and photon to axion (inverse-)decays. Our analysis confirmed that while the annihilations are subdominant and hence negligible in general, inverse decays can significantly contribute to the decaying DM ALP fraction (besides just the Primakoff process), and thus cannot be overlooked in axion cosmology. For ease of future applications, we provide simple fitting formulae for the freeze-in abundance, derived from the corresponding equilibrium abundance computed at characteristic temperatures corresponding to Primakoff and inverse decays. Depending upon their mass and ALP-photon coupling constant, ALPs ultimately decay away into photons once the forward decay rate becomes comparable to the Hubble parameter. For the case when this decay happens prior to recombination, we also computed the resulting increment of the plasma temperature, and the accompanying reduction in the effective number of neutrino species, $\Delta N_{\rm eff}$. 

We considered ALP to ALP annihilations. For the parameter space studied, this gives rise to a negligible abundance of photophobic ALPs, even in the case that the parent species comes into thermal equilibrium. This is due to the quartic coupling being $\lambda \approx m_\phi^2/\mathcal{F}^2$ with $m_\phi\ll \mathcal{F}$ in the range studied. However, in string theory models such as Ref.~\cite{Gendler:2023kjt}, there can exist ALPs with masses all the way up to the Kaluza-Klein scale, for which $\lambda\sim\mathcal{O}(1)$. ALP to ALP annhilaitons give a significant abundance for $\lambda\gtrsim 10^{-10}$, which for $\mathcal{F}\sim 10^{13}\text{ GeV}$ occurs for $m_\phi\gtrsim 10^8\text{ GeV}$. If the reheating temperature is large enough to produce such heavy ALPs, then ALP to ALP annihilations may become important. We will study this further in future work.

Finally, we introduced an effective temperature for the non-thermal ALPs produced via freeze-in, and compared it with equivalent thermal relics of the same mass that decouple at $T_{\rm reh}$. We found that the former is typically hotter than the latter by $20\% - 80\%$ (depending on ALP mass and reheating temperature). This last finding could have implications for constraints on ALPs derived from late time decays in the Milky Way. For example, in the ``irreducible'' axion scenario with $T_{\rm reh}=T_{\rm BBN}$~\cite{Langhoff:2022bij}, the most powerful limits at $m_a=\mathcal{O}(1) \text{ keV}$ are derived from X-ray spectra, and the absence of a decaying DM line therein~\cite{Foster:2021ngm}. The limits depend on the density profile of the ALPs in the galactic centre, which is assumed to follow the density profile of thermal warm DM of the same particle mass. In the case that ALPs are hotter than the equivalent thermal relic, as we have found, they will have a more extended profile, possibly affecting the derived constraints.

We presented fitting parameters for the freeze-in axion abundance, $A_{\rm Prim}$ and $A_{\rm id}$, which may be useful for future studies of freeze-in production of axions. We note that Ref.~\cite{Gendler:2023kjt} used a value of $A_{\rm Prim}=1/0.16=6.25$. This was found at higher reheating temperatures than we considered in this work, $T_{\rm reh}\gtrsim 1$ GeV, by using the fitting formulae in Ref.~\cite{Balazs:2022tjl} and compared to the analytic result using the equilibrium abundance and the Primakoff rate for massless particles from Ref.~\cite{Depta:2020wmr}. While $A_{\rm Prim}$ indeed increases at high values of $T_{\rm reh}$ as reported and discussed in this work, it is unclear whether it increases to $\sim 6$ at high reheating temperatures (beyond QCD crossover), when other charged species including quarks should also be taken into consideration. This is because such massive charged species were not taken into account in the fitting formula of Ref.~\cite{Balazs:2022tjl}. Future studies of freeze-in axions in the axiverse should make use of the improved calculations presented here, while the calculations presented in this work will be extended in future to higher values of $T_{\rm reh}$ where more charged degrees of freedom become active.

We emphasize that in the calculations carried out in this work, we have adopted some approximations pertaining to plasma effects. In particular, we have considered an effective photon mass (equal to the plasma frequency) only in the photon propagator in the Primakoff process, and in the external photon states of the inverse decay. However, in order to accurately account for the plasma effects, one must compute finite temperature photon and fermion resumed propagators, dispersion relations, spectral functions, etc. We shall carry out a full computation with these finite temperature effects taken into account, in an upcoming work.

In this work, we have considered only the production of ALP particles through freeze-in. The epoch prior to reheating cannot necessarily dilute all of the ALP condensate created unavoidably by vacuum misalignment or topological defects. The post-reheating physics of a classical field, namely ``the condensate", can however be modelled.\footnote{See Ref.~\cite{Eberhardt:2023axk} and references therein for discussion on quantum corrections to the classical coherent state evolution, and Ref.~\cite{Marsh:2022gnf} for discussion on measuring the classical state in the laboratory.} In Ref.~\cite{Balazs:2022tjl}, the condensate was accounted for simplistically by allowing the relative axion DM abundance $\xi$ to be a free parameter. Interactions between ALP fields in two-field misalignment have been studied in Refs.~\cite{Cyncynates:2021xzw,Cyncynates:2022wlq} while non-linear interactions between the condensate and gauge fields have been investigated in Ref.~\cite{Anzuini:2024rpl}. On the other hand, oscillating classical fields experience dissipation, typically featuring particle production. In a plasma, the dissipation process is more complicated than the zero-temperature case as there could be scattering processes between the particles in the plasma and the oscillating condensate~\cite{Ai:2021gtg,Ai:2023ahr}. In Ref.~\cite{Ai:2023qnr}, using non-equilibrium quantum field theory, some of us developed a formalism to study interactions between the ALP condensate and freeze-in ALPs. Further detailed study of the condensate and condensate-particle interactions will be the subject of future work.

\acknowledgments

The work of WYA is supported by the UK Engineering and Physical Sciences Research Council (EPSRC), under Research Grant No. EP/V002821/1. DJEM is supported by an Ernest Rutherford Fellowship from the Science and Technologies Facilities Council (ST/T004037/1). DJEM and MJ are supported by a Leverhulme Trust Research Project (RPG-2022-145). We are grateful to Jens Chluba, Sebastian Hoof, Marco Hufnagel, Felix Kahlhoefer, Georg  Raffelt and Nicholas Rodd for the discussions. 

\newpage

\appendix

\section{Calculating collision terms}\label{appendixA}

In this section, we derive the production rates for ALPs from all the three processes described in Fig.~\ref{fig:Feyn}. The interaction Hamiltonian is
\begin{align}
\label{eq:H_int}
    H_{\rm int} = \int\mathrm{d}^3x\Bigl[\frac{g_{\phi\gamma}}{4}\phi\tilde{F}F + e\bar{\psi}\slashed{A}\psi\Bigr],
\end{align}
where $\psi$ is the fermion field and $\phi$ is the ALP field. Also, $g = 1/f_a$ is the ALP-photon coupling. With a finite volume V, the fields can be Fourier decomposed as usual:
\begin{align}
    \phi(x) &= \sum_{\bm k}\frac{1}{\sqrt{2V\omega^{\phi}_{\bm k}}}\Bigl[a_{\bm k}\,\e^{-i(\omega^{\phi}_{\bm k}t - {\bm k}\cdot{\bm x})} \,+\, a^{\dagger}_{\bm k}\,\e^{i(\omega^{\phi}_{\bm k}t - {\bm k}\cdot{\bm x})}\Bigr]\,,\nonumber\\
    A^{\mu}(x) &= \sum_{\bm k, \lambda}\frac{1}{\sqrt{2V\omega^{\gamma}_{\bm k,\lambda}}}\Bigl[a_{\bm k, \lambda}\,\varepsilon^{\mu}_{\bm k,\lambda}\,\e^{-i(\omega^{\gamma}_{\bm k,\lambda}t - {\bm k}\cdot{\bm x})} \,+\, a^{\dagger}_{\bm k, \lambda}\,\varepsilon^{\mu\,\ast}_{\bm k,\lambda}\,\e^{i(\omega^{\gamma}_{\bm k,\lambda}t - {\bm k}\cdot{\bm x})}\Bigr]\,,\nonumber\\
    \psi(x) &= \sum_{\bm k, s}\frac{1}{\sqrt{2V\omega^{e}_{\bm k}}}\Bigl[a_{\bm k, s}\,u_{\bm k, s}\,\e^{-i(\omega^{e}_{\bm k}t - {\bm k}\cdot{\bm x})} \,+\, b^{\dagger}_{\bm k, s}\,v_{\bm k, s}\,\e^{i(\omega^{e}_{\bm k}t - {\bm k}\cdot{\bm x})}\Bigr]\,,
\end{align}
where $\omega^{x}_{\bm k} = \sqrt{m_x^2 + {\bm k}^2}$ is the free dispersion relation for electron and axion. For both photons and electrons, we take their distributions to be the thermal equilibrium distributions throughout this work:
\begin{align}
    f^x_{\bm k} = \frac{1}{{\rm e}^{\omega^{x}_{\bm k}/T} \pm 1}\qquad x=\gamma\,\ ({\rm with} -1)\;;\;x=e\ ({\rm with} +1)\;.
\end{align}
Here and in the following, we will use non-bold letters to denote four-coordinates and four-momenta.\\ 

In order to \textit{roughly} capture plasma effects, we keep the photon mass $m_{\gamma}$ only in (1) the photon propagator for the Primakoff process to handle the co-linear collision logarithmic divergence (as usually done for long-range interactions); (2) the external photon line for the inverse decay, to retain a threshold. This is the same approach as taken in Ref.~\cite{Cadamuro:2010cz}. We take the photon mass to be given by the following
\begin{align}
\label{eq:photonmass}
    m^2_{\gamma} \equiv \frac{e^2n_e}{\langle\omega^{e}\rangle}\,,
\end{align} 
where $n_e$ and $\langle\omega^e\rangle$ are the equilibrium number density of electrons (including positrons), and the average energy of the electron respectively.

We would like to mention that while the above approximations lead to a simpler analysis, a complete treatment requires one to compute finite temperature corrected propagators and spectral functions for the various plasma species. Such finite temperature effects may play an important role, as shown for example in some astrophysical settings~\cite{Chanda:1987ax,Elmfors:1997tt,Drewes:2021fjx,Li:2022dkc,Bouzoud:2024bom}. We will present this analysis in an upcoming paper.

\subsection{Primakoff process}
\label{app:PartPrim_rate}

The tree level contribution to the scattering matrix for $\phi_{k} + e_{p,s} \rightarrow e_{q,s'} + \gamma_{\ell, \lambda}$ (where the subscript denotes the respective particle's $4$ momentum and spin helicity) is 
\begin{align}
    i\mathcal{M} &= \frac{g_{\phi\gamma}e}{4}\langle \gamma_{\ell,\lambda} e_{q,s'}|{\rm T}\Bigl[\int\mathrm{d}^4x\int\mathrm{d}^4x'\,\bar{\psi}(x')\gamma_{\sigma}\psi(x')\,A^{\sigma}(x')\,\phi(x)\,\tilde{F}_{\mu\nu}(x)F^{\mu\nu}(x)\Bigr]|\phi_{k} e_{p,s}\rangle\,,
\end{align}
which after using the previous Fourier decomposition along with usual manipulations, gives
\begin{align}
    i\mathcal{M} &= \frac{-ig_{\phi\gamma}e\,(2\pi)\delta(\omega^{\gamma}_{\bm \ell} - \omega^{\phi}_{\bm k} - \omega^{e}_{\bm p} + \omega^{e}_{\bm q})\,\delta_{{\bm \ell} - {\bm k}, {\bm p} - {\bm q}}}{V\sqrt{2\omega^e_{\bm q}}\,\sqrt{2\omega^e_{\bm p}}\,\sqrt{2\omega^{\phi}_{\bm k}}\,\sqrt{2\omega^{\gamma}_{\bm \ell}}}\,\frac{1}{((p-q)^2 - m_{\gamma}^2)}\,\times\nonumber\\  
    &\qquad\qquad\qquad\qquad\qquad\qquad\qquad\qquad\qquad\,\epsilon_{\mu\nu\alpha\beta}\,\bar{u}_{\bm q, s'}\gamma^{\nu}u_{\bm p, s}\ell^{\alpha}\varepsilon^{\beta}_{\ell,\lambda}(p^{\mu}-q^{\mu})\,.
\end{align}
To handle the Dirac delta formally, we can discretize time together with $(2\pi)\delta(\omega_1-\omega_2) \rightarrow T\delta_{\omega_1,\omega_2}$. The modulus square of the above matrix element divided by time, that is the probability rate $\mathcal{P}_r \equiv |\mathcal{M}|^2/T$ of this reaction, is then
\begin{align}
    \mathcal{P}_r &= \frac{g^2_{\phi\gamma}e^2\,(2\pi)\delta(\omega^{\gamma}_{\bm \ell} - \omega^{\phi}_{\bm k} - \omega^{e}_{\bm p} + \omega^{e}_{\bm q})\,\delta_{{\bm \ell} - {\bm k}, {\bm p} - {\bm q}}}{V^2(2\omega^e_{\bm q})(2\omega^e_{\bm p})(2\omega^{\phi}_{\bm k})(2\omega^{\gamma}_{\bm \ell})}\frac{1}{((p-q)^2-m_{\gamma}^2)^2}\,\times\nonumber\\
    &\qquad\qquad{\rm Tr}\Bigl\{\bar{u}_{\bm q, s'}\gamma^{\nu}u_{\bm p, s}\bar{u}_{\bm p, s}\gamma^{\nu'}u_{\bm q,s'}\Bigr\}\epsilon_{\mu\nu\alpha\beta}\,\epsilon_{\mu'\nu'\alpha'\beta'}\ell^{\alpha}\varepsilon^{\beta}_{\bm \ell, \lambda}(p^{\mu}-q^{\mu})\,\ell^{\alpha'}\varepsilon^{\beta'\,\ast}_{\bm \ell, \lambda}(p^{\mu'}-q^{\mu'})\,.
\end{align}
Here in the prefactor, we first discarded the redundant Kronecker delta, and then took the $T \rightarrow \infty$ limit once again.

We shall now assume unpolarized initial and final states, and hence sum over all polarizations $s$, $s'$, and $\lambda$. Using the identity $\sum_{s}u_{\bm k, s}\bar{u}_{\bm k, s} = \slashed{k}+m_e$, for the trace we get ${\rm Tr}\Bigl\{(\slashed{q}+m_e)\gamma^{\nu}(\slashed{p}+m_e)\gamma^{\nu'}\Bigr\} = {\rm Tr}\Bigl\{\slashed{q}\gamma^{\nu}\slashed{p}\gamma^{\nu'} + m_e^2\gamma^{\nu}\gamma^{\nu'}\Bigr\} = 4(q^{\nu}p^{\nu'} + q^{\nu'}p^{\nu} - \eta^{\nu\nu'}(p\cdot q - m_e^2))$. Similarly, there is another ``identity" for the photon polarization, $\sum_{\lambda}\epsilon^{\beta'\,\ast}_{\bm \ell, \lambda}\epsilon^{\beta}_{\bm \ell, \lambda} \rightarrow -\eta^{\beta\beta'}$, that we use.\footnote{Not surprisingly, the $\ell^{\mu}\ell^{\sigma}/m_{\gamma}^2$ bit in the polarization summation replacement (if the photon were to be truly massive containing $3$ independent degrees of freedom) does not contribute. This is reflective of the preservation of gauge invariance (relatable to the Ward identity).} Furthermore, we use the identity
$\varepsilon^{\nu\mu\alpha\beta}\varepsilon_{\nu\mu'\alpha'\beta'} = \delta^{\mu}_{\mu'}(\delta^{\alpha}_{\alpha'}\delta^{\beta}_{\beta'} - \delta^{\alpha}_{\beta'}\delta^{\beta}_{\alpha'}) + \delta^{\mu}_{\alpha'}(\delta^{\alpha}_{\beta'}\delta^{\beta}_{\mu'} - \delta^{\alpha}_{\mu'}\delta^{\beta}_{\beta'}) + \delta^{\mu}_{\beta'}(\delta^{\alpha}_{\mu'}\delta^{\beta}_{\alpha'} - \delta^{\alpha}_{\alpha'}\delta^{\beta}_{\mu'})$, and then also $\ell^2 = 0$ and $p^2 = q^2 = m_e^2$ (on account of on-shell photon in the final state and electron in the initial and final state respectively). Finally, we attach the appropriate occupation number functions (to account for the non-zero populations of particles already in the plasma and also to account for both the forward and backward reactions). 

After all of this, we obtain 
\begin{align}
    \sum_{s,s',\lambda}\mathcal{P}_r\ast &= \frac{(2\pi)\delta(\omega^{\gamma}_{\bm \ell} - \omega^{\phi}_{\bm k} - \omega^{e}_{\bm p} + \omega^{e}_{\bm q})\,\delta_{{\bm \ell} - {\bm k}, {\bm p} - {\bm q}}}{V^2(2\omega^e_{\bm q})(2\omega^e_{\bm p})(2\omega^{\phi}_{\bm k})(2\omega^{\gamma}_{\bm \ell})}\frac{4\pi \alpha g^2_{\phi\gamma}}{((p-q)^2-m_{\gamma}^2)^2}\Biggl[8(p\cdot q - 2m_e^2)(\ell\cdot(p-q))^2\nonumber\\
    &\, - 8(p-q)^2(\ell\cdot q)(\ell\cdot p)\Biggr]\Bigl[(1+f^{\phi}_{\bm k})(1-f^{e}_{\bm p})f^{\gamma}_{\bm \ell}\,f^{e}_{\bm q} - f^{\phi}_{\bm k}\,f^{e}_{\bm p}\,(1+f^{\gamma}_{\bm \ell})(1-f^{e}_{\bm q})\Bigr]\,,
\end{align}
where $f^{x}_{\bm k}$ is the occupation number function for the specie $x$ (as a function of its momentum ${\bm k}$). Now to calculate the full reaction rate, we would also require a summation over ${\bm p}$, ${\bm q}$, and ${\bm \ell}$, and further take the large volume limit. For this purpose, we send $V^{-1}\sum_{\bm p} \rightarrow (2\pi)^{-3}\int\mathrm{d}^3{\bm p}$ together with $V\delta_{\bm k, \bm k'} \rightarrow (2\pi)^{3}\delta^3({\bm k}-{\bm k}')$. In order to compare it with the literature~\cite{Cadamuro:2010cz}, we re-write everything in terms of Mandelstam variables; $s = (p+k)^2 = (\ell+q)^2$, $t = (\ell-k)^2 = (p-q)^2$, and $u = (k-q)^2 = (\ell-p)^2 = m_{\phi}^2 + 2m_e^2 - s - t$ ; after which we get for $\sum_{\bm p, \bm q, \bm \ell}\sum_{s,s',\lambda}\mathcal{P}\ast  \xrightarrow{V\rightarrow\infty} \mathcal{R}_{\rm Prim}$ the following:
\begin{align}
\label{eq:R_Prim}
    &\mathcal{R}_{\rm Prim} =  \int\frac{\mathrm{d}^3{\bm p}}{(2\pi)^3}\frac{\mathrm{d}^3{\bm q}}{(2\pi)^3}\frac{\mathrm{d}^3\bm\ell}{(2\pi)^3}\frac{(2\pi)^4\,\delta^3({\bm \ell - \bm k - \bm p + \bm q})\,\delta(\omega^{\gamma}_{\bm \ell} - \omega^{\phi}_{\bm k} - \omega^{e}_{\bm p} + \omega^{e}_{\bm q})}{(2\omega^e_{\bm q})(2\omega^e_{\bm p})(2\omega^{\phi}_{\bm k})(2\omega^{\gamma}_{\bm \ell})}\,\times\nonumber\\ 
    &\qquad \frac{4\pi\alpha g^2_{\phi\gamma}}{(t^2-m_{\gamma}^2)^2}\Biggl[-2m_e^2m_{\phi}^4 - 2t^2(s-m_{\phi}^2) - t^3 - t\Bigl(m_{\phi}^4+2(s-m_e^2)^2 - 2m_{\phi}^2(s+m_e^2)\Bigr)\Biggr]\nonumber\\
    &\qquad\qquad\qquad\times\Bigl[(1+f^{\phi}_{\bm k})(1-f^{e}_{\bm p})f^{\gamma}_{\bm \ell}\,f^{e}_{\bm q} - f^{\phi}_{\bm k}\,f^{e}_{\bm p}\,(1+f^{\gamma}_{\bm \ell})(1-f^{e}_{\bm q})\Bigr]\,.
\end{align}
This matches with Eq.~(A.5) of Ref.~\cite{Cadamuro:2010cz}.

Finally, we can identify the two forward and backward pieces as mentioned in the main text (Eq.\eqref{eq:fboltzman}):
\begin{align}
\label{eq:collision_PartPrim}
    \mathcal{C}_{\rm Prim}({\bm k}) &\equiv \int\frac{\mathrm{d}^3{\bm p}}{(2\pi)^3}\frac{\mathrm{d}^3{\bm q}}{(2\pi)^3}\frac{\mathrm{d}^3 \bm\ell}{(2\pi)^3}\frac{(2\pi)^4\,\delta^3({\bm \ell - \bm k - \bm p + \bm q})\,\delta(\omega^{\gamma}_{\bm \ell} - \omega^{\phi}_{\bm k} - \omega^{e}_{\bm p} + \omega^{e}_{\bm q})}{(2\omega^e_{\bm q})(2\omega^e_{\bm p})(2\omega^{\phi}_{\bm k})(2\omega^{\gamma}_{\bm \ell})}\,\times\nonumber\\ 
    &\qquad \frac{4\pi\alpha g^2_{\phi\gamma}}{(t - m_{\gamma}^2)^2}\Biggl[-2m_e^2m_{\phi}^4 - 2t^2(s-m_{\phi}^2) - t^3 - t\Bigl(m_{\phi}^4+2(s-m_e^2)^2 - 2m_{\phi}^2(s+m_e^2)\Bigr)\Biggr]\nonumber\\
    &\qquad\qquad\qquad\times\Bigl[(1-f^{e}_{\bm p})f^{\gamma}_{\bm \ell}\,f^{e}_{\bm q}\Bigr]\nonumber\\
    \mathcal{C}'_{\rm Prim}({\bm k}) &\equiv \int\frac{\mathrm{d}^3{\bm p}}{(2\pi)^3}\frac{\mathrm{d}^3{\bm q}}{(2\pi)^3}\frac{\mathrm{d}^3\bm\ell}{(2\pi)^3}\frac{(2\pi)^4\,\delta^3({\bm \ell - \bm k - \bm p + \bm q})\,\delta(\omega^{\gamma}_{\bm \ell} - \omega^{\phi}_{\bm k} - \omega^{e}_{\bm p} + \omega^{e}_{\bm q})}{(2\omega^e_{\bm q})(2\omega^e_{\bm p})(2\omega^{\phi}_{\bm k})(2\omega^{\gamma}_{\bm \ell})}\,\times\nonumber\\ 
    &\qquad \frac{4\pi\alpha g^2_{\phi\gamma}}{(t - m_{\gamma}^2)^2}\Biggl[-2m_e^2m_{\phi}^4 - 2t^2(s-m_{\phi}^2) - t^3 - t\Bigl(m_{\phi}^4+2(s-m_e^2)^2 - 2m_{\phi}^2(s+m_e^2)\Bigr)\Biggr]\nonumber\\
    &\qquad\qquad\qquad\times\Bigl[f^{e}_{\bm p}(1+f^{\gamma}_{\bm \ell})(1-f^{e}_{\bm q})\Bigr]
\end{align}

\subsubsection{Simplifying the integrals for numerical calculation}

We begin by simply integrating out ${\bm p}$ (by virtue of the momentum conserving Dirac delta), and also redefine ${\bm \ell} = {\bm Q}+{\bm k}$. We then integrate over the c-angle between ${\bm q}$ and ${\bm Q}$. Letting $x\equiv \vec{Q}\cdot {\bm q}/(|\vec{Q}||{\bm q}|)$, the zero of the Dirac delta gives the following root for $x$ (which must be restricted to lie within the unit mod interval):
\begin{align}
\label{eq:root_xr}
   -1 \leq x_r \equiv \frac{(\omega^e_{\bm q} + \omega^{\gamma}_{\bm k + \bm Q} -\omega^{\phi}_{\bm k})^2 - (\omega^{e}_{\bm q})^2 - {\bm Q}^2}{2|{\bm q}||{\bm Q}|} \leq 1\,.
\end{align}
Defining the vectors ${\bm q}$ and ${\bm k}$ with respect to ${\bm Q}$, lets us integrate the azimuthal angle of ${\bm Q}$ trivially. Finally, with $y\equiv \vec{Q}\cdot{\bm k}/(|\vec{Q}||{\bm k}|)$  and $\varphi$ as the azimuthal angle of ${\bm q}$ (as measured w.r.t. the vector ${\bm Q}$), we note that any term that contains odd powers of $\cos \varphi$ or $\sin \varphi$ will result in zero. Noting that the thermal/Boltzmann occupation number functions of electrons and photons do not depend on any of the azimuthal angles, the only piece that contains $\varphi$ dependent terms would be ${\bm k}\cdot{\bm q} = |{\bm k}||{\bm q}|(\cos\theta\,\cos\theta_{\bm k} + \cos(\varphi-\varphi_{\bm k})\sin\theta\,\sin\theta_{\bm k})$, where $\theta=\arccos x$ and $(\theta_{\bm k}, \varphi_{\bm k})$ are the angles of ${\bm k}$ w.r.t. ${\bm Q}$. After integrating out $\varphi$ and $\varphi_{\bm k}$, we finally get
\begin{align}
    &\mathcal{C}_{\rm Prim}(|{\bm k}|) = \frac{8\alpha g^2_{\phi\gamma}}{2(2\pi)^2}\int_{\rm reg}\frac{\mathrm{d}(|{\bm q}|)|{\bm q}|\,\mathrm{d}(|{\bm Q}|)|{\bm Q}|\,\mathrm{d}y}{(2\omega^e_{\bm q})(2\omega^{\phi}_{\bm k})(2\omega^{\gamma}_{\bm k + \bm Q})\,\left((\omega^{\gamma}_{\bm k + \bm Q} - \omega^{\phi}_{\bm k})^2-{\bm Q}^2-m_{\gamma}^2\right)^2}\times\nonumber\\
    &\; \Biggl[A(|{\bm k}|,|{\bm q}|,|{\bm Q}|,y) - \frac{1}{4{\bm Q^2}}\left(\omega^{\gamma}_{{\bm k}+{\bm Q}} - \omega^{\phi}_{\bm k} + |{\bm Q}|\right)\left(-\omega^{\gamma}_{{\bm k}+{\bm Q}} + \omega^{\phi}_{\bm k} + |{\bm Q}|\right)B(|{\bm k}|,|{\bm q}|,|{\bm Q}|,y)\Biggr]\times\nonumber\\
    &\qquad(1-f^{e}_{\bm q + \bm Q})\,f^{\gamma}_{\bm k + \bm Q}\,f^{e}_{\bm q}\,,\nonumber\\
    &\mathcal{C}'_{\rm Prim}(|{\bm k}|) = \frac{8\alpha g^2_{\phi\gamma}}{2(2\pi)^2}\int_{\rm reg}\frac{\mathrm{d}(|{\bm q}|)|{\bm q}|\,\mathrm{d}(|{\bm Q}|)|{\bm Q}|\,\mathrm{d}y}{(2\omega^e_{\bm q})(2\omega^{\phi}_{\bm k})(2\omega^{\gamma}_{\bm k + \bm Q})\,\left((\omega^{\gamma}_{\bm k + \bm Q} - \omega^{\phi}_{\bm k})^2-{\bm Q}^2-m_{\gamma}^2\right)^2}\times\nonumber\\
    &\;\Biggl[A(|{\bm k}|,|{\bm q}|,|{\bm Q}|,y) - \frac{1}{4{\bm Q^2}}\left(\omega^{\gamma}_{{\bm k}+{\bm Q}} - \omega^{\phi}_{\bm k} + |{\bm Q}|\right)\left(-\omega^{\gamma}_{{\bm k}+{\bm Q}} + \omega^{\phi}_{\bm k} + |{\bm Q}|\right)B(|{\bm k}|,|{\bm q}|,|{\bm Q}|,y)\Biggr]\times\nonumber\\
    &\qquad f^{e}_{\bm q + \bm Q}(1+f^{\gamma}_{\bm k + \bm Q})(1-f^{e}_{\bm q})\,,
\end{align}
where
\begin{align}
    A(|{\bm k}|,|{\bm q}|,|{\bm Q}|,y) &\equiv \left(|{\bm k}||{\bm Q}| y+\omega^{\gamma}_{{\bm k}+{\bm Q}} (\omega^{\phi}_{\bm k} - \omega^{\gamma}_{{\bm k}+{\bm Q}}) + {\bm Q}^2\right)^2\left(-2m_e^2 - (\omega^{\gamma}_{{\bm k}+{\bm Q}} - \omega^{\phi}_{\bm k})^2 + {\bm Q}^2\right)\,,
\end{align}
and
\begin{align}
    &B(|{\bm k}|,|{\bm q}|,|{\bm Q}|,y) \equiv 2 {\bm Q}^2 \left(-(\omega^{\gamma}_{{\bm k}+{\bm Q}} - \omega^{\phi}_{\bm k})^2 + 2\omega^{e}_{\bm q}\omega^{\phi}_{\bm k} + 
   {\bm Q}^2\right)\left(-(\omega^{\gamma}_{{\bm k}+{\bm Q}})^2 - 2\omega^{e}_{\bm q}\omega^{\phi}_{\bm k} + (\omega^{\phi}_{\bm k})^2 + {\bm Q}^2\right)\nonumber\\
   &\qquad\qquad + 4 |{\bm k}||{\bm Q}| y\left(\omega^{\phi}_{\bm k}(\omega^{\gamma}_{{\bm k}+{\bm Q}} - \omega^{\phi}_{\bm k}) (2\omega^{e}_{\bm q} + \omega^{\gamma}_{{\bm k}+{\bm Q}} - \omega^{\phi}_{\bm k})^2 + \omega^{\gamma}_{{\bm k}+{\bm Q}} (-\omega^{\gamma}_{{\bm k}+{\bm Q}} + \omega^{\phi}_{\bm k}) {\bm Q}^2 + {\bm Q}^4\right)\nonumber\\
   &\qquad\qquad + {\bm k}^2 \Bigl[4{\bm q}^2{\bm Q}^2(-1 + y^2) + \left(-((\omega^{\gamma}_{{\bm k}+{\bm Q}} - \omega^{\phi}_{\bm k}) (2\omega^{e}_{\bm q} + \omega^{\gamma}_{{\bm k}+{\bm Q}} - \omega^{\phi}_{\bm k})) + {\bm Q}^2\right)\times\nonumber\\
   &\qquad\qquad\qquad \left({\bm Q}^2 (1 + y^2) + (\omega^{\gamma}_{{\bm k}+{\bm Q}} - \omega^{\phi}_{\bm k}) (2\omega^{e}_{\bm q} + \omega^{\gamma}_{{\bm k}+{\bm Q}} - \omega^{\phi}_{\bm k}) (-1 + 3 y^2)\right)\Bigr].
\end{align}
Also, ``reg" is the region defined by the constraint~\eqref{eq:root_xr}. 

\begin{figure}[ht]\centering\includegraphics[scale=0.25]{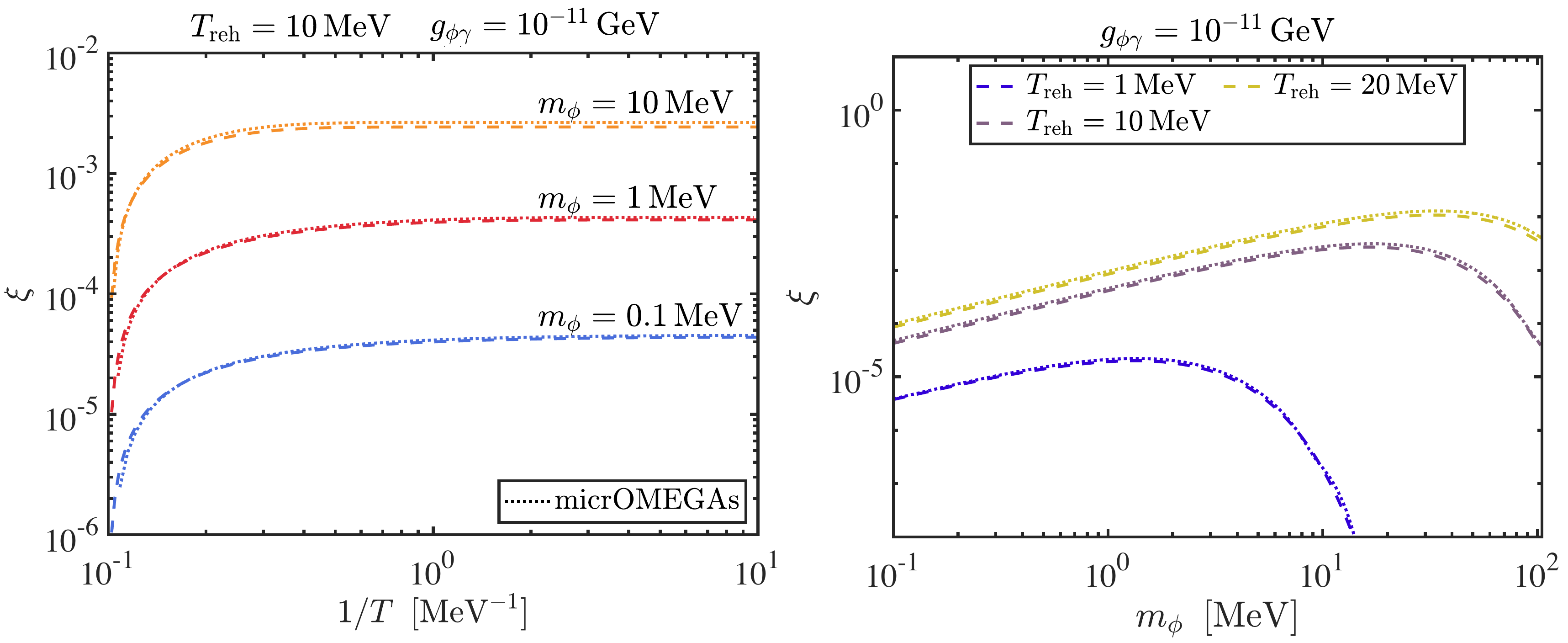}
\caption{Comparison between our calculation and the one using \texttt{micrOMEGAs} v5.3.41 for the Primakoff process only. Left panel: The freeze-in abundance $\xi$ for three different ALP masses at the same $T_{\rm reh}$. Right panel: Comparison of $\xi$ as a function of ALP mass $m_{\phi}$, for three different $T_{\rm reh}$. Note that $\xi$ here is different from that used in the main text. Here we have made the replacement $\rho(t)\rightarrow m_\phi n(t)$ in order to compare with the result of \texttt{micrOMEGAs} v5.3.41.}
\label{fig:xiplot_compare_MOmega}
\end{figure}

We evaluate the above integrals numerically, for various different values of $m_{\phi}$. In the left panel of Fig.~\ref{fig:xiplot_compare_MOmega}, we show a comparison of the ALP fraction $\xi$ (c.f. Eq.~\eqref{eq:xi_def} with $\rho(t) \rightarrow m_{\phi}n(t)$), between our Primakoff calculation and the equivalent result obtained using \texttt{micrOMEGAs} v5.3.41 (for the three prototypical masses $m_{\phi}$ considered in the main text and $T_{\rm reh} = 10$ MeV). Also in the right panel of Fig.~\ref{fig:xiplot_compare_MOmega}, we show $\xi$ obtained as a function of $m_{\phi}$ for three different reheating temperatures. We find excellent agreement between our calculation and \texttt{micrOMEGAs} v5.3.41. For higher reheating temperatures close to muon mass, contributions of muons become relevant.

\subsubsection{Evaluating the forward reaction rate in the relativistic limit}
\label{sec:Primakoff_rel_calculation}

Integrating over ${\bm k}$ as well, while dropping Bose enhancement and Pauli blocking factors along with any mass dependence, we have the following for the forward ``rate density"
\begin{align}
    \int\frac{{\rm d}^3{\bm k}}{(2\pi)^3}\mathcal{C}_{\rm Prim}(|{\bm k}|) &\approx \int\frac{{\rm d}^3{\bm k}}{(2\pi)^3}\frac{\mathrm{d}^3{\bm p}}{(2\pi)^3}\frac{\mathrm{d}^3{\bm q}}{(2\pi)^3}\frac{\mathrm{d}^3 \bm\ell}{(2\pi)^3}\frac{(2\pi)^4\,\delta^3({\bm \ell - \bm k - \bm p + \bm q})\,\delta(|\bm\ell| - |\bm k| - |\bm p| + |\bm q|)}{(2|\bm q|)(2|\bm p|)(2|\bm k|)(2|\bm\ell|)}\,\times\nonumber\\ 
    &\qquad \frac{8\pi\alpha g^2_{\phi\gamma}}{(|\bm p| |\bm q| - {\bm p}\cdot{\bm q})}\Biggl[(|\bm\ell| |\bm  p| - {\bm \ell}\cdot{\bm p})^2 + (|\bm \ell| |\bm q| - {\bm \ell}\cdot{\bm q})^2\Biggr]{\rm e}^{-|\bm \ell|/T}{\rm e}^{-|\bm q|/T}\,.
\end{align}
The energy-momentum conservation enforces $(|\bm p||\bm q| - {\bm p}\cdot{\bm q}) = (|\bm \ell| |\bm q| - {\bm \ell}\cdot{\bm q}) - (|\bm \ell| |\bm p| - {\bm \ell}\cdot{\bm p})$, giving
\begin{align}
    \int\frac{{\rm d}^3{\bm k}}{(2\pi)^3}\mathcal{C}_{\rm Prim}(|{\bm k}|) &\approx \int\frac{{\rm d}^3{\bm k}}{(2\pi)^3}\frac{\mathrm{d}^3{\bm p}}{(2\pi)^3}\frac{\mathrm{d}^3{\bm q}}{(2\pi)^3}\frac{\mathrm{d}^3 \bm\ell}{(2\pi)^3}\frac{(2\pi)^4\,\delta^3({\bm \ell - \bm k - \bm p + \bm q})\,\delta(|\bm \ell| - |\bm k| - |\bm p| + |\bm q|)}{(2|\bm q|)(2|\bm p|)(2|\bm k|)(2|\bm\ell|)}\,\times\nonumber\\ 
    &\qquad 8\pi\alpha g^2_{\phi\gamma}\Biggl[\frac{(|\bm \ell| |\bm p| - {\bm \ell}\cdot{\bm p})^2 + (|\bm \ell| |\bm q| - {\bm \ell}\cdot{\bm q})^2}{(|\bm \ell| |\bm q| - {\bm \ell}\cdot{\bm q}) - (|\bm \ell| |\bm p| - {\bm \ell}\cdot{\bm p})}\Biggr]{\rm e}^{-|\bm \ell|/T}{\rm e}^{-|\bm q|/T}\,.
\end{align}
We can first integrate over ${\bm k}$ trivially (on account of the $3$-momentum conserving Dirac delta). Then, we can define the vectors ${\bm q}$ and ${\bm \ell}$ with respect to ${\bm p}$, which lets us integrate over the solid angle associated with ${\bm p}$ trivially, followed by one azimuthal angle (out of the two associated with ${\bm q}$ and ${\bm \ell}$). Afterwards, we can integrate over $|\bm p|$ by virtue of the remaining (energy-conserving) Dirac delta. Performing these steps fetches 
\begin{align}
    &\int\frac{{\rm d}^3{\bm k}}{(2\pi)^3}\mathcal{C}_{\rm Prim}(|{\bm k}|) \approx \frac{\alpha g^2_{\phi\gamma}}{2}\int\frac{\mathrm{d} |\bm q|\,|\bm q|}{2\pi^2}\frac{{\rm d}y}{2}\frac{{\rm d}\varphi}{2\pi}\frac{\mathrm{d}|\bm \ell|\,|\bm \ell|^2}{2\pi^2}\frac{{\rm d}x}{2}\frac{{\rm e}^{-|\bm \ell|/T}{\rm e}^{-|\bm q|/T}}{\left(|\bm \ell|(1-x)+|\bm q|(1-y)\right)^3}\,\times\nonumber\\ 
    &\qquad \frac{|\bm \ell| |\bm q|}{(1-y)}\Biggl[\left(1 - xy - \cos\varphi\sqrt{(1-x^2)(1-y^2)}\right)^2\times\nonumber\\
    &\qquad\qquad\qquad\qquad\qquad\left(2 |\bm \ell|^2 (1-x)^2 + 2|\bm \ell| |\bm q|(1-x) (1-y) + |\bm q|^2(1-y)^2\right)\Biggr]\,.
\end{align}
where $x$ and $y$ are the c-angles between ${\bm \ell}$ and ${\bm p}$, and ${\bm q}$ and ${\bm p}$ respectively. Integration over $\varphi$ is trivial. In order to proceed, we can massage the denominator into exponent by using the identity $B^{-3} = T^{-3}\int^{\infty}_{0}{\rm d}z\,z^2\,{\rm e}^{-z B/T}/2!$ where $B = |\bm \ell|(1 - x) + |\bm q|(1 - y)$. We get
\begin{align}
    &\int\frac{{\rm d}^3{\bm k}}{(2\pi)^3}\mathcal{C}_{\rm Prim}(|{\bm k}|) \approx \frac{\alpha g^2_{\phi\gamma}}{2T^3}\int\frac{{\rm d}z\,z^2}{2}\frac{\mathrm{d}|\bm q|\,|\bm q|^2}{2\pi^2}\frac{{\rm d}y}{2}\frac{\mathrm{d}|\bm \ell|\,|\bm\ell|^3}{2\pi^2}\frac{{\rm d}x}{2}\,{\rm e}^{-\frac{|\bm\ell|}{T}[1+z(1-x)]}{\rm e}^{-\frac{|\bm q|}{T}[1+z(1-y)]}\,\times\nonumber\\ 
    &\qquad \frac{1}{(1-y)}\Biggl[\left((1 - xy)^2 + \frac{(1-x^2)(1-y^2)}{2}\right)\times\nonumber\\
    &\qquad\qquad\qquad\qquad\qquad\qquad\left(2 |\bm \ell|^2 (1-x)^2 + 2|\bm \ell| |\bm q|(1-x) (1-y) + |\bm q|^2(1-y)^2\right)\Biggr]\,.
\end{align}
Integrating out $|\bm \ell|$ and $|\bm q|$ gives
\begin{align}
    &\int\frac{{\rm d}^3{\bm k}}{(2\pi)^3}\mathcal{C}_{\rm Prim}(|{\bm k}|) \approx \frac{6\alpha g^2_{\phi\gamma}T^6}{\pi^4}\int\frac{{\rm d}z\,z^2}{2}\frac{{\rm d}y}{2}\frac{{\rm d}x}{2}\left(\frac{1}{(1+z(1-x))^6\,(1+z(1-y))^5}\right)\,\times\nonumber\\ 
    &\qquad \frac{(-1)}{(1-y)}\Biggl[\left((1 - xy)^2 + \frac{(1-x^2)(1-y^2)}{2}\right)\times\nonumber\\
    &\left(-10 x^2-19 (x-1)^2 (y-1)^2 z^2+2 (x-1) (y-1) z (13 x+6 y-19)-6 x y+26 x-3 y^2+12 y-19\right)\Biggr]\,.
\end{align}
Integrating out $z$ and $x$, and also redefining $y \rightarrow y-1$ gives
\begin{align}
    &\int\frac{{\rm d}^3{\bm k}}{(2\pi)^3}\mathcal{C}_{\rm Prim}(|{\bm k}|) \approx \frac{4\alpha g^2_{\phi\gamma}T^6}{\pi^4}\int^2_0\frac{{\rm d}y}{2}\frac{1}{(y-2)y^5}\,\times\nonumber\\ 
    &\qquad \Biggl[y \left(144+72\log(32)-y \left(3 y^2-41 y+((y-22) y+148) \log (2)+144\right)\right) \nonumber\\
    &\qquad\qquad + (y-12)(y-6)(y-2)^2\log(2-y) - 288\log(2)\Biggr]\,.
\end{align}
The above integral has a log divergence as $y \rightarrow 2$ (which is the collinear collision limit). Supplying the cutoff $y_{\rm cutoff}=2- 2m^2_{\gamma}/s$, we get
\begin{align}
    \int\frac{{\rm d}^3{\bm k}}{(2\pi)^3}\mathcal{C}_{\rm Prim}(|{\bm k}|) \approx \frac{\alpha g^2_{\phi\gamma}T^6}{2\pi^4}\Bigl[2\log\left(\frac{\sqrt{s}}{m_{\gamma}}\right) - \frac{5}{4}\Bigr]\,. 
\end{align}
Here the Mandelstam variable, $s = (\langle\omega^e\rangle + \langle\omega^{\gamma}\rangle)^2 \approx 34.25T^2$, is the square of the center of mass energy. Finally using $n_{\rm eq} = \zeta(3)T^3/\pi^2$ (and also including an extra factor of $2$ to include the same process but with positrons), this gives the following rate
\begin{align}
    \Gamma_{\rm eq, \rm Prim} \equiv \frac{1}{n_{\rm eq}}\int\frac{{\rm d}^3{\bm k}}{(2\pi)^3}\mathcal{C}_{\rm Prim}(|{\bm k}|) \approx  \frac{\alpha g^2_{\phi\gamma}T^3}{\pi^2\zeta(3)}\Bigl[2\log\left(\frac{T}{m_{\gamma}}\right) + 2.28\Bigr]\,.
\end{align}
We note that the above is not exactly equal to the result from~\cite{Bolz:2000fu}, as used in the main text (c.f. Eq.~\eqref{eq:Gamma_Prim_eq}). This is because in our computation above, we neglected the Bose enhancement and Pauli blocking effects for the photons and fermions respectively. That is, we took $f_{\bm p} \approx {\rm e}^{-|\bm p|/T}$ for both species.

\subsection{Electron-Positron annihilation process}
\label{app:annihilation_rate}

Since this process is just a rotation of the Primakoff process, we can directly obtain its rate using crossing symmetry. That is, sending $\{k,p,q,\ell\} \rightarrow \{k,-p,q,-\ell\}$ together with appropriate changes in the factor containing occupation number functions in Eq.~\eqref{eq:R_Prim}. After these manipulations (along with dropping $m_{\gamma}$ in the photon propagator), we get
\begin{align}
\label{eq:R_ann}
    &\mathcal{R}_{\rm ann} =  \int\frac{\mathrm{d}^3{\bm p}}{(2\pi)^3}\frac{\mathrm{d}^3{\bm q}}{(2\pi)^3}\frac{\mathrm{d}^3\bm\ell}{(2\pi)^3}\frac{(2\pi)^4\,\delta^3({-\bm \ell - \bm k + \bm p + \bm q})\,\delta(-\omega^{\gamma}_{\bm \ell} - \omega^{\phi}_{\bm k} + \omega^{e}_{\bm p} + \omega^{e}_{\bm q})}{(2\omega^e_{\bm q})(2\omega^e_{\bm p})(2\omega^{\phi}_{\bm k})(2\omega^{\gamma}_{\bm \ell})}\,\times\nonumber\\ 
    &\qquad \frac{4\pi\alpha g^2_{\phi\gamma}}{(p+q)^4}\Biggl[8(p\cdot q + 2m_e^2)(\ell\cdot(p+q))^2 - 8(p+q)^2(\ell\cdot q)(\ell\cdot p)\Biggr]\nonumber\\
    &\qquad\qquad\qquad\times\Bigl[(1+f^{\phi}_{\bm k})(1+f^{\gamma}_{\bm \ell})\,f^{e}_{\bm p}\,f^{e}_{\bm q} - f^{\phi}_{\bm k}\,f^{\gamma}_{\bm \ell}\,(1-f^{e}_{\bm p})(1-f^{e}_{\bm q})\Bigr]\,.
\end{align}

\subsubsection{Simplifying the integrals for numerical calculation}

\begin{figure}[h]
    \centering
\includegraphics[width=0.7\linewidth]{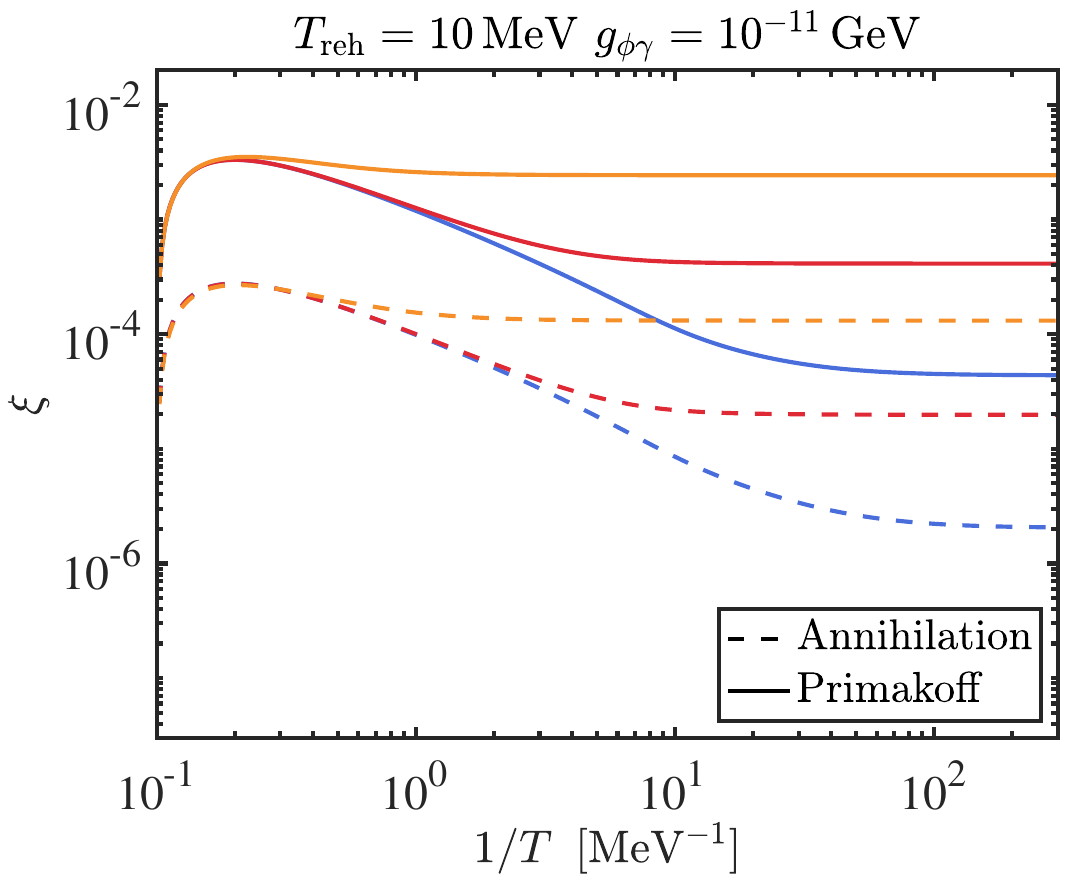}
    \caption{The ALP fraction $\xi$ for our three prototypical ALP masses, due to Primakoff process (sold) and electron-positron annihilation process (dashed). The latter is more than one order of magnitude smaller than the Primakoff, hence we neglect it for our main purposes in this paper.}
\label{fig:annihilation_vs_Primakoff}
\end{figure}

Like in the previous case, we begin by integrating out ${\bm p}$ (by virtue of the momentum conserving Dirac delta), and also redefine ${\bm \ell} = {\bm Q}-{\bm k}$. We then integrate over the c-angle between ${\bm q}$ and ${\bm Q}$. Letting $x\equiv \vec{Q}\cdot {\bm q}/(|\vec{Q}||{\bm q}|)$, the zero of the Dirac delta gives the following root for $x$ (which must be restricted to lie within the unit mod interval):
\begin{align}
\label{eq:root_xr2}
   -1 \leq x_r \equiv \frac{(- \omega^{\gamma}_{\bm Q - \bm k} -\omega^{\phi}_{\bm k} + \omega^e_{\bm q})^2 - (\omega^{e}_{\bm q})^2 - {\bm Q}^2}{2|{\bm q}||{\bm Q}|} \leq 1\,.
\end{align}
Defining the vectors ${\bm q}$ and ${\bm k}$ with respect to ${\bm Q}$, lets us integrate the azimuthal angle of ${\bm Q}$ trivially. Finally, with $y\equiv \vec{Q}\cdot{\bm k}/(|\vec{Q}||{\bm k}|)$  and $\varphi$ as the azimuthal angle of ${\bm q}$ (as measured w.r.t. the vector ${\bm Q}$), we note that any term that contains odd powers of $\cos \varphi$ or $\sin \varphi$ will result in zero. Noting that the thermal/Boltzmann occupation number functions of electrons and photons do not depend on any of the azimuthal angles, the only piece that contains $\varphi$ dependent terms would be ${\bm k}\cdot{\bm q} = |{\bm k}||{\bm q}|(\cos\theta\,\cos\theta_{\bm k} + \cos(\varphi-\varphi_{\bm k})\sin\theta\,\sin\theta_{\bm k})$, where $\theta=\arccos x$ and $(\theta_{\bm k}, \varphi_{\bm k})$ are the angles of ${\bm k}$ w.r.t. ${\bm Q}$. After integrating out $\varphi$ as well, we get
\begin{align}
    &\mathcal{C}_{\rm ann}(|{\bm k}|) = \frac{8\alpha g^2_{\phi\gamma}}{2(2\pi)^2}\int_{\rm reg}\mathrm{d}(|{\bm q}|)|{\bm q}|\,\mathrm{d}(|{\bm Q}|)|{\bm Q}|\,\mathrm{d}y\,\times\nonumber\\
    &\qquad\frac{1}{(2\omega^e_{\bm q})(2\omega^{\phi}_{\bm k})(2\omega^{\gamma}_{\bm Q - \bm k})\left[\left((\omega^{\phi}_{\bm k}+\omega^{\gamma}_{\bm Q-\bm k})^2 - {\bm Q}^2 - m_{\gamma}^2\right)^2\right]}\times\nonumber\\
    &\qquad \Biggl[\tilde{A}(|{\bm k}|,|{\bm q}|,|{\bm Q}|,y) - \frac{1}{4{\bm Q^2}}\left({\bm Q}^2 - (\omega^{\phi}_{\bm k} + \omega^{\gamma}_{\bm Q-\bm k})^2\right)\tilde{B}(|{\bm k}|,|{\bm q}|,|{\bm Q}|,y)\Biggr]\,(1 + f^{\gamma}_{\bm Q - \bm k})\,f^{e}_{\bm Q - \bm q}\,f^{e}_{\bm q}\,,\nonumber\\
    &\mathcal{C}'_{\rm ann}(|{\bm k}|) = \frac{8\alpha g^2_{\phi\gamma}}{2(2\pi)^2}\int_{\rm reg}\mathrm{d}(|{\bm q}|)|{\bm q}|\,\mathrm{d}(|{\bm Q}|)|{\bm Q}|\,\mathrm{d}y\,\times\nonumber\\
    &\qquad\frac{1}{(2\omega^e_{\bm q})(2\omega^{\phi}_{\bm k})(2\omega^{\gamma}_{\bm Q - \bm k})\left[\left((\omega^{\phi}_{\bm k}+\omega^{\gamma}_{\bm Q-\bm k})^2 - {\bm Q}^2 - m_{\gamma}^2\right)^2\right]}\times\nonumber\\
    &\qquad \Biggl[\tilde{A}(|{\bm k}|,|{\bm q}|,|{\bm Q}|,y) - \frac{1}{4{\bm Q^2}}\left({\bm Q}^2 - (\omega^{\phi}_{\bm k} + \omega^{\gamma}_{\bm Q-\bm k})^2\right)\tilde{B}(|{\bm k}|,|{\bm q}|,|{\bm Q}|,y)\Biggr]\times\nonumber\\
    &\qquad f^{\gamma}_{\bm Q - \bm k}\,(1-f^{e}_{\bm Q - \bm q})(1-f^{e}_{\bm q})\,,
\end{align}
where
\begin{align}
    \tilde{A}(|{\bm k}|,|{\bm q}|,|{\bm Q}|,y) &\equiv \left(y|{\bm k}||{\bm Q}| + \omega^{\gamma}_{{\bm Q}-{\bm k}} (\omega^{\phi}_{\bm k} + \omega^{\gamma}_{{\bm Q}-{\bm k}}) - {\bm Q}^2\right)^2\left(2m_e^2 - (\omega^{\gamma}_{{\bm Q}-{\bm k}} + \omega^{\phi}_{\bm k})^2 + {\bm Q}^2\right)\,,
\end{align}
and
\begin{align}
    \tilde{B}(|{\bm k}|,|{\bm q}|,|{\bm Q}|,y) &\equiv -2 {\bm Q}^2 \left({\bm Q}^2 + (\omega^{\phi}_{\bm k})^2 - (\omega^{\gamma}_{\bm Q - \bm k})^2 - 2\omega^{e}_{\bm q}\omega^{\phi}_{\bm k}\right)\left({\bm Q}^2 - (\omega^{\phi}_{\bm k}+\omega^{\gamma}_{{\bm Q}-{\bm k}})^2 + 2\omega^{e}_{\bm q}\omega^{\phi}_{\bm k}\right)\nonumber\\
    &\quad + 4 |{\bm k}||{\bm Q}| y\Bigl({\bm Q}^4 - {\bm Q}^2\omega^{\gamma}_{\bm Q-\bm k}(\omega^{\gamma}_{\bm Q-\bm k} + \omega^{\phi}_{\bm k}) - \omega^{\phi}_{\bm k}(\omega^{\gamma}_{\bm Q-\bm k} + \omega^{\phi}_{\bm k})(\omega^{\gamma}_{\bm Q-\bm k} + \omega^{\phi}_{\bm k} - 2\omega^e_{\bm q})^2\Bigr)\nonumber\\
    &\quad - {\bm k}^2 \Bigl[4{\bm q}^2{\bm Q}^2(-1 + y^2) + \left({\bm Q}^2 - (\omega^{\gamma}_{{\bm Q}-{\bm k}} + \omega^{\phi}_{\bm k}) (\omega^{\gamma}_{{\bm Q}-{\bm k}} + \omega^{\phi}_{\bm k} - 2\omega^{e}_{\bm q})\right)\times\nonumber\\
    &\qquad\qquad\left({\bm Q}^2 (1 + y^2) + (-1 + 3 y^2)(\omega^{\gamma}_{{\bm Q}-{\bm k}} + \omega^{\phi}_{\bm k}) (\omega^{\gamma}_{{\bm Q}-{\bm k}} + \omega^{\phi}_{\bm k} - 2\omega^{e}_{\bm q})\right)\Bigr]\,.
\end{align}
Also, ``reg" is the region defined by the constraint~\eqref{eq:root_xr2}. We evaluate the above integrals numerically, for various different values of $m_{\phi}$. In Fig.~\ref{fig:annihilation_vs_Primakoff}, we show the contributions from the Primakoff process and the annihilation process respectively, towards the ALP fraction $\xi$. The latter contributions are subdominant (by more than a decade), consistent with the findings of Ref.~\cite{Langhoff:2022bij}.

\subsubsection{Evaluating the forward reaction rate in the relativistic limit}
\label{sec:annihilation_rel_calculation}

Integrating over ${\bm k}$ as well, while dropping Bose enhancement and Pauli blocking factors along with any mass dependence, we have the following for the forward ``rate density"
\begin{align}
    \int\frac{{\rm d}^3{\bm k}}{(2\pi)^3}\mathcal{C}_{\rm ann}(|{\bm k}|) &\approx \int\frac{{\rm d}^3{\bm k}}{(2\pi)^3}\frac{\mathrm{d}^3{\bm p}}{(2\pi)^3}\frac{\mathrm{d}^3{\bm q}}{(2\pi)^3}\frac{\mathrm{d}^3 \bm\ell}{(2\pi)^3}\frac{(2\pi)^4\,\delta^3({-\bm \ell - \bm k + \bm p + \bm q})\,\delta(-|\bm \ell| - |\bm k| + |\bm p| + |\bm q|)}{(2|\bm q|)(2|\bm p|)(2|\bm k|)(2|\bm\ell|)}\,\times\nonumber\\ 
    &\qquad \frac{8\pi\alpha g^2_{\phi\gamma}}{(|\bm p| |\bm q| - {\bm p}\cdot{\bm q})}\Biggl[(|\bm \ell| |\bm p| - {\bm \ell}\cdot{\bm p})^2 + (|\bm \ell| |\bm q| - {\bm \ell}\cdot{\bm q})^2\Biggr]{\rm e}^{-|\bm p|/T}{\rm e}^{-|\bm q|/T}\,.
\end{align}
The energy-momentum conservation enforces $(|\bm p| |\bm q| - {\bm p}\cdot{\bm q}) = (|\bm \ell| |\bm p| - {\bm \ell}\cdot{\bm p}) + (|\bm \ell| |\bm q| - {\bm \ell}\cdot{\bm q})$, giving
\begin{align}
    \int\frac{{\rm d}^3{\bm k}}{(2\pi)^3}\mathcal{C}_{\rm ann}(|{\bm k}|) &\approx \int\frac{{\rm d}^3{\bm k}}{(2\pi)^3}\frac{\mathrm{d}^3{\bm p}}{(2\pi)^3}\frac{\mathrm{d}^3{\bm q}}{(2\pi)^3}\frac{\mathrm{d}^3 \bm\ell}{(2\pi)^3}\frac{(2\pi)^4\,\delta^3(-{\bm \ell - \bm k + \bm p + \bm q})\,\delta(-|\bm \ell| - |\bm k| + |\bm p| + |\bm q|)}{(2|\bm q|)(2|\bm p|)(2|\bm k|)(2|\bm\ell|)}\,\times\nonumber\\ 
    &\qquad 8\pi\alpha g^2_{\phi\gamma}\Biggl[\frac{(|\bm \ell| |\bm p| - {\bm \ell}\cdot{\bm p})^2 + (|\bm \ell| |\bm q| - {\bm \ell}\cdot{\bm q})^2}{(|\bm \ell| |\bm p| - {\bm \ell}\cdot{\bm p}) + (|\bm \ell| |\bm q| - {\bm \ell}\cdot{\bm q})}\Biggr]{\rm e}^{-|\bm p|/T}{\rm e}^{-|\bm q|/T}\,.
\end{align}
We can first integrate over ${\bm k}$ trivially (on account of the $3$-momentum conserving Dirac delta). Then, we can define the vectors ${\bm q}$ and ${\bm p}$ with respect to ${\bm \ell}$, which lets us integrate over the solid angle associated with ${\bm \ell}$ trivially, followed by one azimuthal angle (out of the two associated with ${\bm p}$ and ${\bm q}$). Afterwards, we can integrate over $|\bm \ell|$ by virtue of the remaining (energy-conserving) Dirac delta. Performing these steps fetches
\begin{align}
    \int\frac{{\rm d}^3{\bm k}}{(2\pi)^3}\mathcal{C}_{\rm ann}(|{\bm k}|) &\approx \int\frac{\mathrm{d}|\bm p|\,|\bm p|}{2\pi^2}\frac{{\rm d}x}{2}\frac{\mathrm{d}|\bm q|\,|\bm q|}{2\pi^2}\frac{{\rm d}y}{2}\frac{d\varphi}{2\pi}\times\nonumber\\ 
    &\qquad \frac{\alpha g^2_{\phi\gamma}}{2}|\bm \ell|_0^2\Biggl[\frac{|\bm p|^2(1 - x)^2 + |\bm q|^2(1 - y)^2}{(|\bm p|(1 - x) + |\bm q|(1 - y))^2}\Biggr]{\rm e}^{-|\bm p|/T}{\rm e}^{-|\bm q|/T}\,,
\end{align}
where $x$ and $y$ are the c-angles between ${\bm p}$ and ${\bm \ell}$, and ${\bm q}$ and ${\bm \ell}$ respectively, and
\begin{align}
    |\bm \ell|_0 = \frac{|\bm p| |\bm q|(1-xy-\cos\varphi\sqrt{(1-x^2)(1-y^2)})}{(|\bm p|(1-x)+|\bm q|(1-x))}\,.
\end{align}
We can now perform integration over $\varphi$ trivially. Then to proceed further, we can massage the denominator into exponent by using the identity $B^{-4} = T^{-4}\int^{\infty}_{0}{\rm d}z\,z^3\,{\rm e}^{-z B/T}/3!$ where $B = |\bm p|(1 - x) + |\bm q|(1 - y)$, and write
\begin{align}
    \int\frac{{\rm d}^3{\bm k}}{(2\pi)^3}\mathcal{C}_{\rm ann}(|{\bm k}|) &\approx \frac{\alpha g^2_{\phi\gamma}}{2T^4}\int\frac{{\rm d}z\,z^3}{6}\frac{\mathrm{d}|\bm p|\,|\bm p|}{2\pi^2}\frac{{\rm d}x}{2}\frac{\mathrm{d}|\bm q|\,|\bm q|}{2\pi^2}\frac{{\rm d}y}{2}\,\Bigl[|\bm p|^2(1 - x)^2 + |\bm q|^2(1 - y)^2\Bigr]\times\nonumber\\ 
    & |\bm p|^2|\bm q|^2\Bigl((1-xy)^2 + \frac{(1-x^2)(1-y^2)}{2}\Bigr){\rm e}^{-\frac{|\bm p|}{T}[1+z(1-x)]}{\rm e}^{-\frac{|\bm q|}{T}[1+z(1-y)]}\,.
\end{align}
Integration over $|\bm q|$ and $|\bm p|$ gives
\begin{align}
    &\int\frac{{\rm d}^3{\bm k}}{(2\pi)^3}\mathcal{C}_{\rm ann}(|{\bm k}|) \approx \frac{15\alpha g^2_{\phi\gamma}T^6}{4\pi^4}\int{\rm d}z\,z^3\,{\rm d}x\,{\rm d}y\,\Bigl((1-xy)^2 + \frac{(1-x^2)(1-y^2)}{2}\Bigr)\times\nonumber\\ 
    &\qquad \frac{x^2+2 (x-1)^2 (y-1)^2 z^2-2 (x-1) (y-1) z (x+y-2)-2 x+y^2-2 y+2}{(-x z+z+1)^6 (-y z+z+1)^6}\,.
\end{align}
Integration over $x$ and $y$ gives
\begin{align}
    \int\frac{{\rm d}^3{\bm k}}{(2\pi)^3}\mathcal{C}_{\rm ann}(|{\bm k}|) &\approx \frac{160\alpha g^2_{\phi\gamma}T^6}{3\pi^4}\int{\rm d}z\,z^3\frac{1}{(1+2z)^6}\,.
\end{align}
Integration over $z$ gives
\begin{align}
    \int\frac{{\rm d}^3{\bm k}}{(2\pi)^3}\mathcal{C}_{\rm ann}(|{\bm k}|) &\approx \frac{\alpha g^2_{\phi\gamma}T^6}{6\pi^4}\,.
\end{align}
Finally using $n_{\rm eq} = \zeta(3)T^3/\pi^2$, we have the following rate
\begin{align}
    \Gamma_{\rm eq, \rm ann} \equiv \frac{1}{n_{\rm eq}}\int\frac{{\rm d}^3{\bm k}}{(2\pi)^3}\mathcal{C}_{\rm ann}(|{\bm k}|) \approx  \frac{\alpha g^2_{\phi\gamma}T^3}{6\pi^2\zeta(3)}\,.
\end{align}
Similar to the Primakoff case, we remind the reader that here we have approximated the thermal distribution functions as $f_{\bm p} \approx {\rm e}^{-|\bm p|/T}$.

\subsection{ALP-photon decays}
\label{app:dec_rate}

The matrix element for the process $\gamma_{\ell,\lambda} + \gamma_{q,\lambda'} \rightarrow \phi_k$ is
\begin{align}
    i\mathcal{M} &= \frac{g_{\phi\gamma}}{4}\langle \gamma_{\ell,\lambda}\gamma_{q,\lambda'} |{\rm T}\Bigl[\int\mathrm{d}^4x\,\phi(x)\,\tilde{F}_{\mu\nu}(x)F^{\mu\nu}(x)\Bigr]|\phi_{k}\rangle\nonumber\\
    &= -\frac{g_{\phi\gamma}(2\pi)\delta(\omega^{\gamma}_{\bm \ell, \lambda} + \omega^{\gamma}_{\bm q, \lambda'} - \omega^{\phi}_{\bm k})\,\delta_{\bm \ell + \bm q,\bm k}}{\sqrt{V}\sqrt{2\omega^{\phi}_{\bm k}}\,\sqrt{2\omega^{\gamma}_{\bm \ell, \lambda}}\,\sqrt{2\omega^{\gamma}_{\bm q,\lambda'}}}\,\epsilon_{\mu\nu\alpha\beta}\,(\ell^{\mu}\varepsilon^{\nu\,\ast}_{\ell,\lambda})(q^{\alpha}\varepsilon^{\beta\,\ast}_{q,\lambda'})\,.
\end{align}
Its squared modulus divided by time, including summation over polarizations $\lambda$ and $\lambda'$ (assuming a statistically unpolarized bath of photons), gives the probability rate $\mathcal{P}_r \equiv \sum_{\lambda,\lambda'}|\mathcal{M}|^2/T$:
\begin{align}
    \mathcal{P}_r &= -\frac{2g^2_{\phi\gamma}(2\pi)\delta(\omega^{\gamma}_{\bm \ell} + \omega^{\gamma}_{\bm q} - \omega^{\phi}_{\bm k})\,\delta_{\bm \ell + \bm q,\bm k}}{V\,2\omega^{\phi}_{\bm k}\,2\omega^{\gamma}_{\bm \ell}\,2\omega^{\gamma}_{\bm q}}\Bigl((\ell\cdot q)^2 - \ell^2q^2\Bigr)
\end{align}
Like in the previous calculation, we discretized time to handle the Dirac delta (i.e., replaced $2\pi\delta(\omega_1-\omega_2) \rightarrow T\delta_{\omega_1,\omega_2}$, and after squaring and dividing by $T$, made the inverse replacement). Attaching the appropriate occupation number function factors, summing over ${\bm \ell}$ and ${\bm q}$, and taking the large volume limit, gives
\begin{align}
    \mathcal{R}_{\rm decay} &= g^2_{\phi\gamma}\int\frac{\d^3\bm\ell}{(2\pi)^3}\frac{\d^3{\bm q}}{(2\pi)^3}\frac{(2\pi)^4\delta(\omega^{\gamma}_{\bm \ell} + \omega^{\gamma}_{\bm q} - \omega^{\phi}_{\bm k})\delta^3({\bm \ell}+{\bm q}-{\bm k})}{2\omega^{\phi}_{\bm k}\,2\omega^{\gamma}_{\bm \ell}\,2\omega^{\gamma}_{\bm q}}\Bigl((\ell\cdot q)^2-m_{\gamma}^4\Bigr)\times\nonumber\\
    &\qquad\qquad\qquad\qquad\qquad\qquad\qquad\Bigl[f^{\gamma}_{\bm \ell}f^{\gamma}_{\bm q}(1+f^{\phi}_{\bm k}) - f^{\phi}_{\bm k}(1 + f^{\phi}_{\bm \ell})(1 + f^{\phi}_{\bm q})\Bigr]\,.
\end{align}
Here, we assumed a photon mass $m_{\gamma}$ for on-shell (external) photons, i.e. $\ell^2 = q^2 = m_{\gamma}^2$, in order to account for plasma blocking effects. Also, we have attached an overall factor of $1/2$ to account for identical $\gamma$ particles. After integrating out all but one momentum, say ${\bm \ell}$, we get
\begin{align}
    \mathcal{R}_{\rm decay} &= \frac{g^2_{\phi\gamma}m_{\phi}^2\Bigl(m_{\phi}^2 - 4m_{\gamma}^2\Bigr)}{16\pi|{\bm k}|}\int\mathrm{d}|{\bm \ell}|\,|{\bm \ell}|\,\frac{1}{2\omega^{\phi}_{\bm k}\,2\omega^{\gamma}_{\bm \ell}}\Theta(-1<z_0<1)\times\nonumber\\
    &\quad\qquad\qquad\qquad\qquad\qquad \Bigl[f^{\gamma}_{\bm \ell}f^{\gamma}_{\bm k - \bm \ell}(1+f^{\phi}_{\bm k}) - f^{\phi}_{\bm k}(1 + f^{\phi}_{\bm \ell})(1 + f^{\phi}_{\bm k -\bm \ell})\Bigr]_{z = z_0}\,,
\end{align}
where $z$ is the c-angle between ${\bm k}$ and ${\bm \ell}$, restricted to take the value $z_0 = (\omega^{\phi}_{\bm k}\omega^{\gamma}_{\bm \ell}-m_{\phi}^2/2)/|{\bm k}||{\bm \ell}|$ owing to energy conservation. To evaluate the $|{\bm \ell}|$ integral, we need to consider two intervals: $|{\bm k}| \in (0,k_0)$ and $|{\bm k}| \in (k_0,\infty)$, where $k_0 \equiv m_{\phi}\sqrt{(m_{\phi}/2m_{\gamma})^2-1}$. In the first interval, the limits of integration on $|{\bm \ell}|$ are obtained by setting $z_0$ equal to $-1$ and $1$. We get $\ell_{\pm 1} = (k_0\omega^{\phi}_{k}\pm |{\bm k}|\omega^{\phi}_{k_0})/2\omega^{\phi}_{k_0}$ for the two limits. In the second interval, both the limits come from $z_0 = 1$, and the lower limit $\ell_{-1}$ changes sign. The limits therefore go from $-\ell_{-1}$ to $\ell_{+1}$. Gluing these two regimes together, we simply get $|\ell_{-1}|$ and $\ell_{+1}$ as the two limits, giving
\begin{align}
    \mathcal{R}_{\rm decay} &= \frac{g^2_{\phi\gamma}m_{\phi}^2\Bigl(m_{\phi}^2 - 4m_{\gamma}^2\Bigr)}{16\pi|{\bm k}|}\int^{\ell_{+1}}_{|\ell_{-1}|}\frac{\mathrm{d}|{\bm \ell}|\,|{\bm \ell}|}{2\omega^{\phi}_{\bm k}\,2\omega^{\gamma}_{\bm \ell}}\Bigl[f^{\gamma}_{|{\bm \ell}|}f^{\gamma}_{|\bm k-\bm\ell|} - f^{\phi}_{|{\bm k}|}(1+f^{\gamma}_{|{\bm \ell}|}+f^{\gamma}_{|\bm k-\bm\ell|})\Bigr]_{z = z_0}\,.
\end{align}
With a thermal distribution of $\gamma$, i.e. $f^{\gamma}_{\bm q} = (\e^{\omega^{\gamma}_{\bm q}/T}-1)^{-1}$, we finally get
\begin{align}\label{eq:decayrate}
    \mathcal{R}_{\rm decay} &= \frac{g^2_{\phi\gamma}m_{\phi}^2\Bigl(m_{\phi}^2 - 4m_{\gamma}^2\Bigr)T}{64\pi \omega^{\phi}_{\bm k}|{\bm k}|}\log\Biggl(\frac{\sinh((\omega^{\phi}_{\bm k}-\omega^{\gamma}_{|\ell_{-1}|})/2T)\sinh(\omega^{\gamma}_{\ell_{+1}}/2T)}{\sinh(\omega^{\gamma}_{|\ell_{-1}|}/2T)\sinh((\omega^{\phi}_{\bm k}-\omega^{\gamma}_{\ell_{+1}})/2T)}\Biggr)\Bigl[f^{\phi,\,{\rm eq}}_{\bm k}-f^{\phi}_{\bm k}\Bigr]\,,
\end{align}
where $f^{\phi,{\rm eq}}_{\bm k} = (\e^{\omega^{\phi}_{\bm k}/T}-1)^{-1}$. It is to be noted that there is also a Heaviside step function $\Theta(m_\phi^2 - 4m_{\gamma}^2)$ appearing in the above which reflects kinematic plasma blocking, that we suppressed writing. Without the bracketed term $[f^{\phi,\,{\rm eq}}_{\bm k}-f^{\phi}_{\bm k}]$, this is the decay collision term appearing in the main text in Eq.~\eqref{eq:fboltzman}. That is, $\mathcal{R}_{\rm decay} = \mathcal{C}_{\rm decay}[f^{\phi,\,{\rm eq}}_{\bm k}-f^{\phi}_{\bm k}]$. In the limit of $m_{\phi} \gg m_{\gamma}$, we recover the result in the literature~\cite{Cadamuro:2010cz,Depta:2020wmr}:
\begin{align}
    \mathcal{R}_{\rm decay} \approx \frac{g^2_{\phi\gamma}m_{\phi}^4}{64\pi \omega^{\phi}_{\bm k}}\Biggl(1+\frac{2T}{|{\bm k}|}\log\Biggl(\frac{1-\e^{-(\omega^{\phi}_{\bm k}+|{\bm k}|)/2T}}{1-\e^{-(\omega^{\phi}_{\bm k}-|{\bm k}|)/2T}}\Biggr)\Biggr)\Bigl[f^{\phi,\,{\rm eq}}_{\bm k} - f^{\phi}_{\bm k}\Bigr]\,.
\end{align}

\subsection{Quartic interaction $\lambda\phi_1^2\phi_2^2/4$}
\label{ClambdaAppendix}

In this section, we derive the collision term for the process $\phi_1\phi_1\rightarrow \phi_2\phi_2$. Let us denote the momenta for this process as  ${\bm k}$ ($\phi_2$), ${\bm p}$ ($\phi_2$), ${\bm q}$ ($\phi_1$) and ${\bm \ell}$ ($\phi_1$). The momentum label ${\bm k}$ for one of $\phi_2$'s indicates that we are tracing the evolution of $\phi_2$ ALPs. Then the collision term reads
\begin{align}
    \C_\lambda ({\bm k}) =&\lambda^2\int\frac{\d^3 {\bm p}}{(2\pi)^3} \frac{\d^3 {\bm q}}{(2\pi)^3} \frac{\d^3 {\bm \ell}}{(2\pi)^3} \frac{(2\pi)^4 \delta^3({\bm q}+{\bm \ell}-{\bm k}-{\bm p})\delta (\omega^{\phi_2}_{\bm k}+\omega_{{\bm p}}^{\phi_2}-\omega_{{\bm q}}^{\phi_1}-\omega_{{\bm \ell}}^{\phi_1})}{(2\omega_{\bm k}^{\phi_2})(2\omega_{{\bm p}}^{\phi_2}) (2\omega_{{\bm q}}^{\phi_1}) (2\omega_{{\bm \ell}}^{\phi_1})}\notag\\
    &\qquad\qquad\qquad\times \left[(1+f_{\bm k}^{\phi_2}) (1+f_{{\bm p}}^{\phi_2}) f^{\phi_1}_{{\bm q}} f^{\phi_1}_{{\bm \ell}} - f_{{\bm k}}^{\phi_2} f_{{\bm p}}^{\phi_2} (1+f^{\phi_1}_{{\bm q}})(1+f^{\phi_1}_{{\bm \ell}})\right]\,.
\end{align}
Similar to the other collision terms, one can first integrate over, e.g. ${\bm p}$, using the first Dirac delta function. Letting ${\bm \ell}={\bm Q}+{\bm k}$, we then have 
\begin{align}
    \C_\lambda ({\bm k}) =&\lambda^2\int\frac{\d^3 {\bm q}}{(2\pi)^3} \frac{\d^3 {\bm Q}}{(2\pi)^3}\frac{(2\pi) \delta (\omega^{\phi_2}_{\bm k}+\omega_{{\bm Q}+{\bm q}}^{\phi_2}-\omega_{{\bm q}}^{\phi_1}-\omega_{{\bm Q}+{\bm k}}^{\phi_1})}{(2\omega_{\bm k}^{\phi_2})(2\omega_{{\bm Q}+{\bm q}}^{\phi_2}) (2\omega_{{\bm q}}^{\phi_1}) (2\omega_{{\bm Q}+{\bm k}}^{\phi_1})}\notag\\
    &\qquad\times \left[(1+f_{\bm k}^{\phi_2}) (1+f_{{\bm Q}+{\bm q}}^{\phi_2}) f^{\phi_1}_{{\bm q}} f^{\phi_1}_{{\bm Q}+{\bm k}} - f_{{\bm k}}^{\phi_2} f_{{\bm Q}+{\bm q}}^{\phi_2} (1+f^{\phi_1}_{{\bm q}})(1+f^{\phi_1}_{{\bm Q}+{\bm k}})\right]\,.
\end{align}
Denoting $x\equiv {\bm Q}\cdot {\bm q}/(|{\bm Q}||{\bm q}|)$, we can integrate out $x$ by using the remaining Dirac delta function, which gives a constraint for the remaining integrals:
\begin{align}
    -1 \leq  x_r\equiv \frac{\left(\omega_{{\bm q}}^{\phi_1}+\omega_{{\bm Q}+{\bm k}}^{\phi_1}-\omega_{{\bm k}}^{\phi_2}\right)^2-\left(\omega_{{\bm q}}^{\phi_2}\right)^2-{\bm Q}^2  }{2|{\bm Q}||{\bm q}|} \leq 1 \,.
\end{align}

Before proceeding further, we note that in the situation we consider the distribution functions for both $\phi_1$ and $\phi_2$ are much smaller than unity, i.e. $(1+f)\approx 1$. The remaining integrals over ${\bm q}$ and ${\bm Q}$ can be carried out as discussed earlier. Introducing $y\equiv {\bm Q}\cdot{\bm k}/(|{\bm Q}||{\bm k}|)$, we arrive at 
\begin{align}
\label{eq:C-lambda-exp}
    \C_{\lambda}(|{\bm k}|)= &
    \frac{\lambda^2}{2} \int_{\rm reg} \frac{\d |{\bm Q}|\,\d  |{\bm q}| \,\d y}{(2\pi)^3}  \frac{|{\bm Q}||{\bm q}|}{(2\omega^{\phi_2}_{|{\bm k}|})(2\omega^{\phi_1}_{|{\bm q}|})(2\omega^{\phi_1}_{\sqrt{|{\bm Q}|^2+|{\bm k}|^2+2|{\bm Q}|{\bm k}|y}})}\notag\\
    &\times\left.\Bigl[f_{\sqrt{|{\bm Q}|^2+|{\bm k}|^2+2|{\bm Q}||{\bm k}|y}}^{\phi_1} f_{|{\bm q}|}^{\phi_1}-f_{|{\bm k}|}^{\phi_2} f_{\sqrt{|{\bm Q}|^2+|{\bm q}|^2+2|{\bm Q}||{\bm q}|x}}^{\phi_2}\Bigl]\right|_{x=x_r}\,.
\end{align}

\section{Numerical evolution of the distribution function}
\label{app:numerical_evolvefk}

With $k \equiv |{\bm k}|$, here we briefly provide details of our numerical technique to evolve the distribution function $f_{k}(t) = f(k,t)$, as dictated by Eq.~\eqref{eq:fboltzman}. Writing the physical momentum in terms of comoving momentum, $k = \tilde{k}a_{\ast}/a$, the left-hand side is nothing but the full convective derivative. Here $a_{\ast}$ is the initial time where we normalize the scale factor to unity. In general, the Boltzmann equation for any target species' distribution function takes the following form
\begin{align}
    \frac{\d}{\d t}f\left(\tilde{k}a_{\ast}/a,t\right) = \mathcal{C}\left[\tilde{k}a_{\ast}/a,t,f^s\right]\,.
\end{align}
Here $f^s$ corresponds to all the distribution functions appearing in the collision term. In order to sample the collision terms properly, we work with the log of inverse temperature $x \equiv \log(\beta) = \log(m/T)$ as the time variable, along with the log of comoving momentum $y \equiv \log\tilde{k}$. That is, the above equation takes the form
\begin{align}
    \frac{\d}{\d x}f(y + x_{\ast} + g(x) - x, x) = h(x)\,\mathcal{C}\left[y,x,f^s\right]\,.
\end{align}
Here, $h(x) = \d t/\d x$, and $g(x) = \log(g_s(x)/g_s(x_{\ast}))/3$. In order to obtain the above form, we used entropy conservation to write $a_{\ast}/a = (\beta_{\ast}/\beta)(g_{s,\ast}/g_s)^{1/3}$. This is the equation that we solve numerically. For simplicity, we use the simple Runge-Kutta method and the discrete evolution takes the following form
\begin{align}
    f(y + x_{\ast} + g(x_i)-x_i, x_i) = f(y + x_{\ast} + g(x_{i-1})-x_{i-1}, x_{i-1}) + \epsilon\,h(x_{i-1})\,\mathcal{C}\left[y,x_{i-1},f^s_{i-1}\right]\,,
\end{align}
where $\epsilon$ is the discretization parameter for the `time' variable $x$, and the subscripts $i$ and $i-1$ correspond to current and previous `time' steps. Also, $f^s_{i-1}$ on the right-hand side corresponds to the distribution function(s)  evaluated at the previous time step. That is, $f^s_{i-1} \equiv f^s(y + x_{\ast} + g(x_{i-1})-x_{i-1}, x_{i-1})$. Notice how the first argument (corresponding to the physical momentum) changes from the right-hand side of the above equation to its left-hand side. This is nothing but the red-shifting of physical momentum. Thus at every step of the iteration, we (1) calculate the right-hand side, (2) extend the array of $y$ values by $g(x_i) - (x_{i-1}) - x_i + x_{i-1}$ (towards smaller numbers), (3) add new larger $y$ values in order to maintain the same largest $y$ value as before, (4) assign the value computed on the right-hand side to a new $f$ (on the left-hand side) that has support over the new $y$ array, (5) repeat. Using the distribution function, we can then compute all the required moments/quantities in the end.

\section{Effect of ALP decays on the plasma temperature}
\label{app:reheating_postdecays}

Here we compute the change in plasma temperature owing to the decay of ALPs into photons. This is relevant for as long as the decays happen while the plasma is strongly coupled, meaning the newly produced photons could come into equilibrium with the background plasma.\footnote{On the other hand if they decay after photon decoupling (CMB), we get a secondary population of photons, alongside the usual CMB population, giving rise to CMB spectral distortions.}

\begin{figure}[h]
    \centering
    \includegraphics[width=1\linewidth]{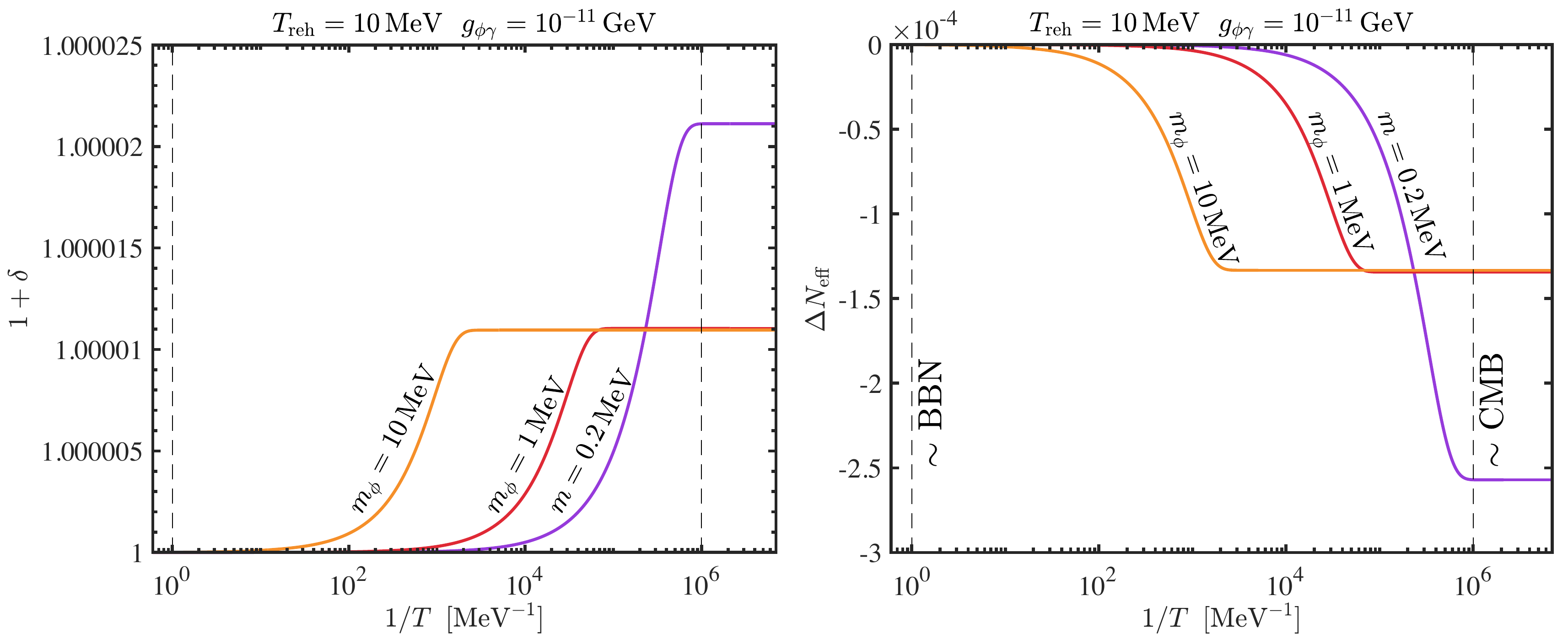}
    \caption{Left Panel: Fractional change in plasma temperature due to ALP decays. The ALPs decay into photons and raise the temperature of the plasma ever so slightly, as expected. Right Panel: The corresponding change in $N_{\rm eff}$ due to entropy injection (c.f. Eq.~\eqref{eq:Delta_Neff}).}
    \label{change_temperature}
\end{figure}

In general, for a plasma in equilibrium, we have the following {\it comoving} entropy
\begin{align}
    s = \frac{a^3}{T}\left(\rho + P - \sum_i\mu_i n_i\right) = a^3\left(g_sT^3 - \sum_i\mu_ig_{i}T^2\right)\,.
\end{align}
Here $a$ is the scale factor, $\rho$ and $P$ are the energy density and pressure of the plasma. The $n_i$ is the number density of the $i^{\rm th}$ species in the plasma, having chemical potential $\mu_i$. The quantity $g_s$ is the effective number of total ``entropic" relativistic degrees of freedom, $g_i$ is the number of relativistic degrees of freedom of the $i^{\rm th}$ species in the plasma, and finally, $T$ is the plasma temperature. In the usual case when there is no entropy ejection (injection) from (into) the plasma, the above stays constant giving rise to the usual temperature evolution $T \sim 1/a$. However, in our case, the production of ALPs (and later their possible depletion back into the photons) leads to non-conservation of this entropy, reflected as a deviation in the temperature's usual evolution. We can compute this change by deriving an equation for the temperature $T$. The change in plasma entropy can be attributed to the change in $\rho$ and $P$ of the photons, which is equal to the negative of the change in energy density and pressure of the ALPs. Furthermore, since the production of ALPs through the Primakoff and inverse decay processes occur at a much lower rate than $H$, the only significant contribution to the change in the plasma temperature from its usual $T \sim 1/a$ behaviour would come from ALP forward decays to photons when $\Gamma_{\rm decay} \gtrsim H$. This causes a heating of the plasma, and we have
\begin{align}
    \d s = -\frac{a^3}{T}\left(\d\rho_\phi + \d P_\phi\right)\,.
\end{align}
Here we can further neglect the pressure term since the ALPs have already become non-relativistic by the time they start to decay. Using $s = (2\pi^2/45)a^3g_sT^3$, we therefore have
\begin{align}
    \frac{2\pi^2}{45}\frac{\d(a^3g_sT^3)}{\d t} = -\frac{a^3}{T}\frac{\d\rho_\phi}{\d t}\,.
\end{align}
Assuming that the change in the plasma temperature is small, i.e. $T = \bar{T}(1 + \delta)$ where the overhead bar corresponds to the usual/unperturbed quantities, the above gives the following leading-order evolution equation for $\delta$ in terms of $\bar{\beta} = m_\phi/\bar{T}$:
\begin{align}
    \frac{\d\delta}{\d\log\bar{\beta}} = -\frac{15}{2\pi^2}\frac{\bar{\beta}^4}{m_\phi^4\,\bar{g}_{s}}\frac{\d\rho_\phi}{\d\log\bar{\beta}}\,.
\end{align}

Using the above, we can further estimate $\Delta N_{\rm eff}$. Writing the full radiation energy density as the sum of photons and neutrinos, $\rho_{\rm rad} = \rho_{\gamma}+ \rho_{\nu}$, where $\rho_{\gamma} = 2(\pi^2/30)T^4$ and $\rho_{\nu} = 2\times(7/8)(\pi^2/30)\times 3 T_\nu^4$, $N_{\rm eff}$ is defined as $\rho_{\rm rad}=(\pi^2/15)[1+(7/8) N_{\rm eff}(4/11)^{4/3}]T^4$. Therefore, we get
\begin{align}
\label{eq:Delta_Neff}
    \Delta N_{\rm eff} = \bar{N}_{\rm eff}\left(\frac{1}{(1+\delta)^4}-1\right) \approx -12.18\,\delta\, \quad {\rm for}\ \delta\ll 1 \,.
\end{align}
Here $\bar{N}_{\rm eff} \approx 3.044$~\cite{Akita:2020szl,Froustey:2020mcq,Bennett:2020zkv,Drewes:2024wbw} (for the usual cosmology without any ALP-induced entropy). This is shown in the right panel of Fig.~\ref{change_temperature}.

\section{Table for $A_{\rm Prim}$}
\label{app:TableAPrim}

\begin{table}[h!]
    \centering
    \caption{Values of $A_{\rm Prim}$ for various $m_{\phi}$ and $T_{\rm reh}$.}
    \vspace{0.5em}
    \renewcommand{\arraystretch}{1.2} 
    \begin{tabular}{|c|c|cccccc|}
        \hline
        \multirow{5}{*}{\raisebox{-6.5em}{\rotatebox[origin=c]{90}{$m_{\phi}$ [MeV]}}} & & \multicolumn{6}{c|}{$T_{\rm reh}$ [MeV]} \\
        \cline{3-8}
        &  & \textbf{5} & \textbf{10} & \textbf{20} & \textbf{30} & \textbf{40} & \textbf{50} \\
        \hline
        & \textbf{0.01} & 1.00 & 1.02 & 1.09 & 1.21 & 1.31 & 1.36 \\
        & \textbf{0.10} & 1.00 & 1.01 & 1.08 & 1.21 & 1.30 & 1.36 \\
        & \textbf{1.00} & 0.91 & 0.96 & 1.05 & 1.11 & 1.28 & 1.34 \\
        & \textbf{10} & 0.67 & 0.76 & 0.90 & 1.04 & 1.15 & 1.22 \\
        & \textbf{100} & 1.73 & 1.07 & 0.81 & 0.80 & 0.82 & 0.84 \\
        \hline
    \end{tabular}
\end{table}

\bibliographystyle{JHEP}
\bibliography{reference}

\end{document}